\documentclass[aps,prl,superscriptaddress,twocolumn]{revtex4-2}

\usepackage{amsmath,amsthm,amssymb,mathrsfs,xcolor,bm,graphicx,dcolumn,array,ulem,soul}

\newcommand{\ie}{{\it i.e.\ }}

\newcommand{\LSNOthird}{\text{La$_{5/3}$Sr$_{1/3}$NiO$_4$}}
\newcommand{\LSNOacro}{\text{LSNO-$1/3$}}

\newcommand{\LSCOx}{\text{La$_{2-x}$Sr$_x$CuO$_4$}}
\newcommand{\LSCOacro}{\text{LSCO}}
\newcommand{\LSCOthousandth}{\text{La$_{1.999}$Sr$_{0.001}$CuO$_4$}}

\begin{document}

\title{
Thermal Hall Resistivity as a Unifying Description of Phonon Thermal Hall Effect in Various Insulators
}

\author{Dirui~Wu}
\author{Ying~Kit~Tsui}
\author{Xavier~Loh}
\affiliation{Division of Physics and Applied Physics, School of Physical and Mathematical Sciences, Nanyang Technological University, Singapore 637371, Singapore}
\author{Minseong~Lee}
\affiliation{National High Magnetic Field Laboratory, Los Alamos National Laboratory, Los Alamos, New Mexico 87545, USA}
\author{Eun~Sang~Choi}
\affiliation{National High Magnetic Field Laboratory, Tallahassee, Florida 32310, USA}
\author{Takao~Sasagawa}
\affiliation{Materials and Structures Laboratory, Institute of Science Tokyo, Yokohama, Kanagawa 226-8501, Japan}
\author{Toshimitsu~Ito}
\affiliation{National Institute of Advanced Industrial Science and Technology, Tsukuba, Ibaraki 305-8565, Japan}
\author{Xiao~Renshaw~Wang}
\affiliation{Division of Physics and Applied Physics, School of Physical and Mathematical Sciences, Nanyang Technological University, Singapore 637371, Singapore}
\affiliation{School of Electrical and Electronic Engineering, Nanyang Technological University, Singapore 639798, Singapore}
\author{Pinaki~Sengupta}
\author{Xinyang~Zhang}
\email[]{xinyang.zhang@ntu.edu.sg}
\author{Christos Panagopoulos}
\email[]{christos@ntu.edu.sg}
\affiliation{Division of Physics and Applied Physics, School of Physical and Mathematical Sciences, Nanyang Technological University, Singapore 637371, Singapore}

\date{\today}

\begin{abstract}
A considerable phonon thermal Hall effect was recently discovered across a diverse collection of materials. To clarify this enigmatic thermal Hall response in various insulators, we investigate the doped Mott insulator \LSNOthird\ as an example material system and reveal a characteristic phonon-dominated thermal Hall effect. As a sensitive probe of the transverse thermal response, the thermal Hall resistivity $w_{xy}$ exhibits an insulating-like temperature dependence $w_{xy}(T)$, a linear magnetic-field dependence $w_{xy}(H)$ near $H=0$, and a $T$-linear thermal Hall angle at low temperatures. The presence of similar phenomena across a series of insulators suggests that $w_{xy}$ serves as a unifying description of phonon thermal Hall effect, corroborated by an apparent correlation between the insulating-like $w_{xy}(T)$ and material's localized electronic state.
\end{abstract}

\maketitle


The thermal Hall effect (THE), \ie the transverse thermal response to a heat current in a time-reversal-symmetry-breaking field, appears to be ubiquitous in materials. Direct electron THE was well established in metals and alloys \cite{behnia_fundamentals_2015}. Recently, THE in insulating quantum magnets emerged as a sensitive probe of chirality in charge-neutral magnetic quasiparticles \cite{onose_observation_2010, hirschberger_large_2015, hirschberger_thermal_2015, watanabe_emergence_2016, kasahara_unusual_2018, akazawa_topological_2022}, typically as a consequence of non-trivial magnon band topology \cite{katsura_theory_2010, murakami_thermal_2017, sun_negative_2021, zhang_thermal_2024}. In various magnetic or non-magnetic insulators, THE can also be attributed to phonon's coupling with lattice, spin, or charge degrees of freedom \cite{strohm_phenomenological_2005, ideue_giant_2017, sugii_thermal_2017, grissonnanche_giant_2019, li_phonon_2020, boulanger_thermal_2020, grissonnanche_chiral_2020, zhang_anomalous_2021, boulanger_thermal_2022, chen_large_2022, lefrancois_evidence_2022, uehara_phonon_2022, li_phonon_2023, ataei_phonon_2024, chen_planar_2024, sharma_phonon_2024, sharma_phonon_2024-1, xu_thermal_2024, sharma_microscopic_2026, boulanger_thermal_2026}, the proposed mechanism of which was typically specific to the particular material system.

Remarkably, the latest discovery of a large phonon thermal Hall conductivity in a broad collection of insulating or semiconducting materials has stimulated extensive investigations into its enigmatic origin \cite{grissonnanche_giant_2019, li_phonon_2020, boulanger_thermal_2020, jin_discovery_2025}. Certain general features emerge: (a) magnetic-field-linear thermal Hall conductivity $\kappa_{xy}\propto H$ as $H\to0$ \cite{grissonnanche_giant_2019, li_phonon_2020}, (b) relatively large thermal Hall angle $\kappa_{xy}/\kappa_{xx}\sim 10^{-3}$ bounded by an intermediate-temperature peak \cite{grissonnanche_giant_2019, boulanger_thermal_2020, li_phonon_2020, li_phonon_2023, behnia_phonon_2025}, and (c) scaling of $|\kappa_{xy}|/\kappa_{xx}H$ \cite{li_phonon_2023, guo_interaction_2026} or $|\kappa_{xy}|/\kappa_{xx}^2$ \cite{jin_discovery_2025}. In cuprate (doped) Mott insulators, the insensitivity of THE to c-axis anisotropy \cite{grissonnanche_chiral_2020} together with a substantial planar THE \cite{chen_planar_2024} have verified its phononic nature, while in insulating SrTiO$_3$ (STO) and semiconducting silicon, the temperature dependence of $\kappa_{xy}$ qualitatively resembles that of Mott insulators \cite{li_phonon_2020, jin_discovery_2025}. A myriad of explanations have since been proposed, ranging from phonon's coupling with lattice distortion \cite{chen_enhanced_2020}, spin \cite{zhang_thermal_2019, oh_phonon_2025}, or impurities \cite{sun_large_2022, flebus_charged_2022, guo_resonant_2022, mangeolle_phonon_2022}, to an intrinsic source of net phonon chirality \cite{qin_energy_2011, saito_berry_2019, behnia_phonon_2025}. Despite substantial efforts involving advanced thermometry and meticulous modeling, it remains unclear whether a single physical quantity can coherently organize the large phonon thermal Hall responses observed across a broad range of insulating materials.

 In analyzing THE, it is crucial to distinguish the inherent transverse (Hall) response from any material-specific longitudinal magneto-thermal contributions. The thermal Hall \textit{resistivity} $w_{xy}$, which directly measures the transverse temperature gradient generated by a longitudinal heat current, is a sensitive probe of the chirality of phonon motion and is less dependent on longitudinal phonon scattering \cite{guo_resonant_2022, mangeolle_phonon_2022, lyu_phonons_2023}. It is distinct from the commonly presented $\kappa_{xy}=-w_{xy}\kappa_{xx}^2$, whose temperature and magnetic-field dependencies may incorporate longitudinal contributions. Indeed, the magnitude of $w_{xy}/H$ was found consistently within $10^{-7}-10^{-5}\ \text{K}{\cdot}\text{m}{\cdot}\text{W}^{-1}{\cdot}\text{T}^{-1}$ as one unifying feature of phonon THE in various insulators and semiconductors \cite{guo_interaction_2026, ling_phonon_2026}. By scrutinizing $w_{xy}$, including its magnitude and temperature or field dependency, comparative studies across various insulators that exhibit canonical phonon THE may shed light on a unifying phenomenological understanding of phonon THE.

In this Letter, we present a systematic study of THE in the doped Mott insulator \LSNOthird\ (\LSNOacro), in comparison with the iso-structural \LSCOx\ (\LSCOacro), which hosts distinct electronic and magnetic ground states, among various other insulators that demonstrate phonon THE. The thermal Hall resistivity tensor $\mathbf{w}$ is shown to differentiate a phonon THE in the transverse channel from the longitudinal magneto-thermal effect, whereas the thermal Hall conductivity $\kappa_{xy}$ is correlated with the longitudinal thermal conductivity $\kappa_{xx}$. Three characteristic phenomena in the temperature and magnetic-field dependence of the thermal Hall resistivity $w_{xy}$, as well as in the thermal Hall angle, are experimentally identified to coherently characterize the phonon THE in LSNO, LSCO, STO, and silicon. Our finding suggests that $w_{xy}$ serves as a unifying description of phonon THE in a broad collection of insulators, corroborated by an apparent correlation between material's localized electronic state and the thermal Hall resistivity.


Single crystal \LSNOthird\ was grown by the floating-zone method and has been characterized in detail \cite{supp}. \LSNOacro\ and \LSCOacro\ share the same tetragonal crystal structure and thus similar phonon modes, but they differ in spin excitations \cite{takeda_crystal_1990}. At various doping, \LSCOacro\ exhibits a rich phase diagram while \LSNOacro\ remains an insulator \cite{cava_magnetic_1991}. Meanwhile, both systems host charge/spin-ordered stripes in a static or fluctuating form, which in \LSNOacro\ is stabilized by commensuration \cite{chen_charge_1993, zaanen_freezing_1994, tranquada_charge_1998, zachar_landau_1998, yoshizawa_stripe_2000, anissimova_direct_2014}. THE was measured in an applied magnetic field along the $c$-axis of \LSNOacro\ and a heat current along the $a$-axis. Care was taken to properly anneal silver-paste thermal contacts for an Ohmic contact to the NiO$_2$ planes and a reduced thermal boundary resistance \cite{tee_two_2017}. Using \textit{in situ} calibrated Cernox thermometers, we performed dedicated one-heater-three-thermometer thermal Hall experiments to measure the thermal resistivity tensor $\mathbf{w}$ \footnote{The thermal Hall resistivity $w_{xy}$ derives directly from measured quantities: transverse temperature difference, heating power, and sample thickness. The measured $w_{xy}$ was anti-symmetrized as $w_{xy}=(w_{xy}(+H)-w_{xy}(-H))/2$ during data analysis.} and calculate the thermal conductivity tensor $\bm{\kappa}$, under the assumption of an isotropic and field-anti-symmetric $\mathbf{w}$ with a small Hall angle \cite{supp}.


\begin{figure}[!t]
	\centering
	\includegraphics[width=0.85\columnwidth]{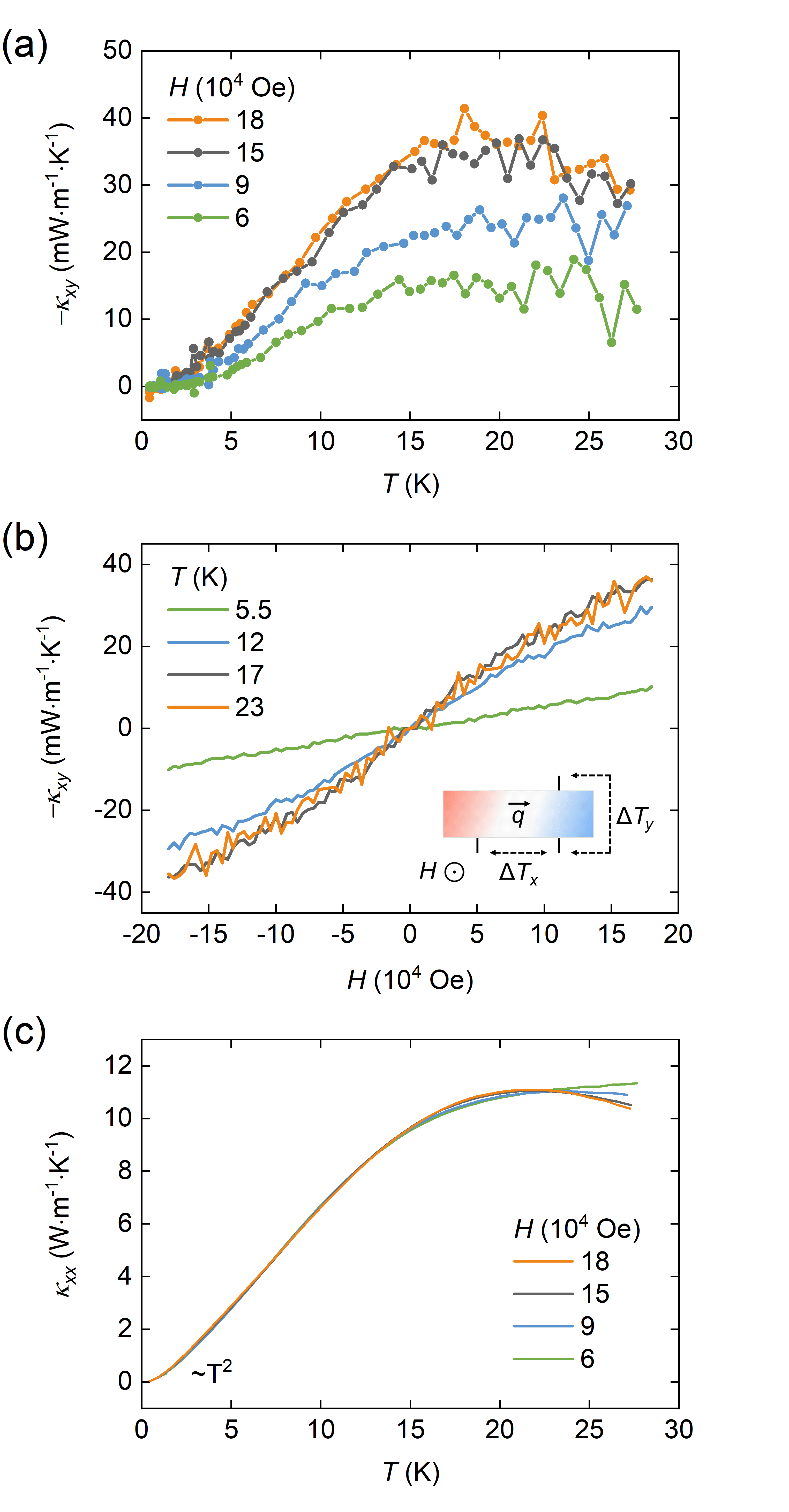}
	\caption{\label{Fig1} Measured thermal conductivity tensor of \LSNOacro. (a)~Temperature dependence of the anti-symmetrized thermal Hall conductivity $-\kappa_{xy}$ at various magnetic field, showing peak features at $\sim20$ K. (b)~Linear or slightly sublinear magnetic-field dependence of $-\kappa_{xy}$ at different temperatures. Inset shows a schematic of the THE experiment with a heat current $\mathbf{q}$ along the $a$-axis of \LSNOacro\ and a magnetic field $H$ along the $c$-axis. (c)~Temperature dependence of the longitudinal thermal conductivity $\kappa_{xx}$ at various magnetic field, following a $T^2$ power-law at low temperatures and exhibiting similar peak features as $\kappa_{xy}$.}
\end{figure}

Figure~\ref{Fig1}(a) shows the thermal Hall conductivity $\kappa_{xy}$ versus temperature between 0.4--28 K in different magnetic field. $\kappa_{xy}(T)$ approximately follows a $T^3$ power-law below 4 K \cite{supp}, before it rises to an intermediate-temperature peak at $T\approx 20$ K (see supplemental material \cite{supp} for higher-temperature THE measurements showing the full peak feature). Figure~\ref{Fig1}(b) shows the magnetic-field dependence of $\kappa_{xy}$ that is linear near $H=0$, albeit turning sub-linear at a higher field. Such temperature and magnetic-field dependencies are consistent with previously reported THE in (doped) Mott insulators \cite{grissonnanche_giant_2019, boulanger_thermal_2020} and other insulating materials in which phonon dominates the thermal transport \cite{li_phonon_2020, jin_discovery_2025}.

Figure~\ref{Fig1}(c) shows peak features of the longitudinal thermal conductivity $\kappa_{xx}$ that strongly resemble those in $\kappa_{xy}$, while below 2 K, $\kappa_{xx}(T)$ follows a $T^2$ power-law \cite{supp} as related to a $T$-independent boundary or defect scattering as the specific heat $c\propto T^2$ for $T\lesssim10$ K in LSNO \cite{matsushita_electrical_1990}. The magnetic-field-induced splitting of $\kappa_{xx}$ for $T\gtrsim20$ K shows a magneto-thermal effect possibly related to spin-lattice coupling \cite{pocs_heat_2025}. As the intermediate-temperature peaks in $\kappa_{xx}$ and $\kappa_{xy}$ are attributed to maximal phonon normal scattering as a signature of phonon thermal transport \cite{behnia_phonon_2025}, we identify a strong correlation between the $\kappa_{xx}$ and $\kappa_{xy}$ of \LSNOacro. This correlation can be justified since the thermal Hall conductivity was calculated as $\kappa_{xy}=-w_{xy}/w_{xx}^2$ in analyzing any experimental THE data. Therefore, the influence of $w_{xx}(T,H)$ has been covertly embedded in Fig.~\ref{Fig1}(a), which may confuse the interpretation of $\kappa_{xy}$ if $w_{xx}$, or equivalently $\kappa_{xx}$, carries a substantial temperature or magnetic-field dependence \cite{mangeolle_phonon_2022}.

\begin{figure}[!t]
	\centering
	\includegraphics[width=0.85\columnwidth]{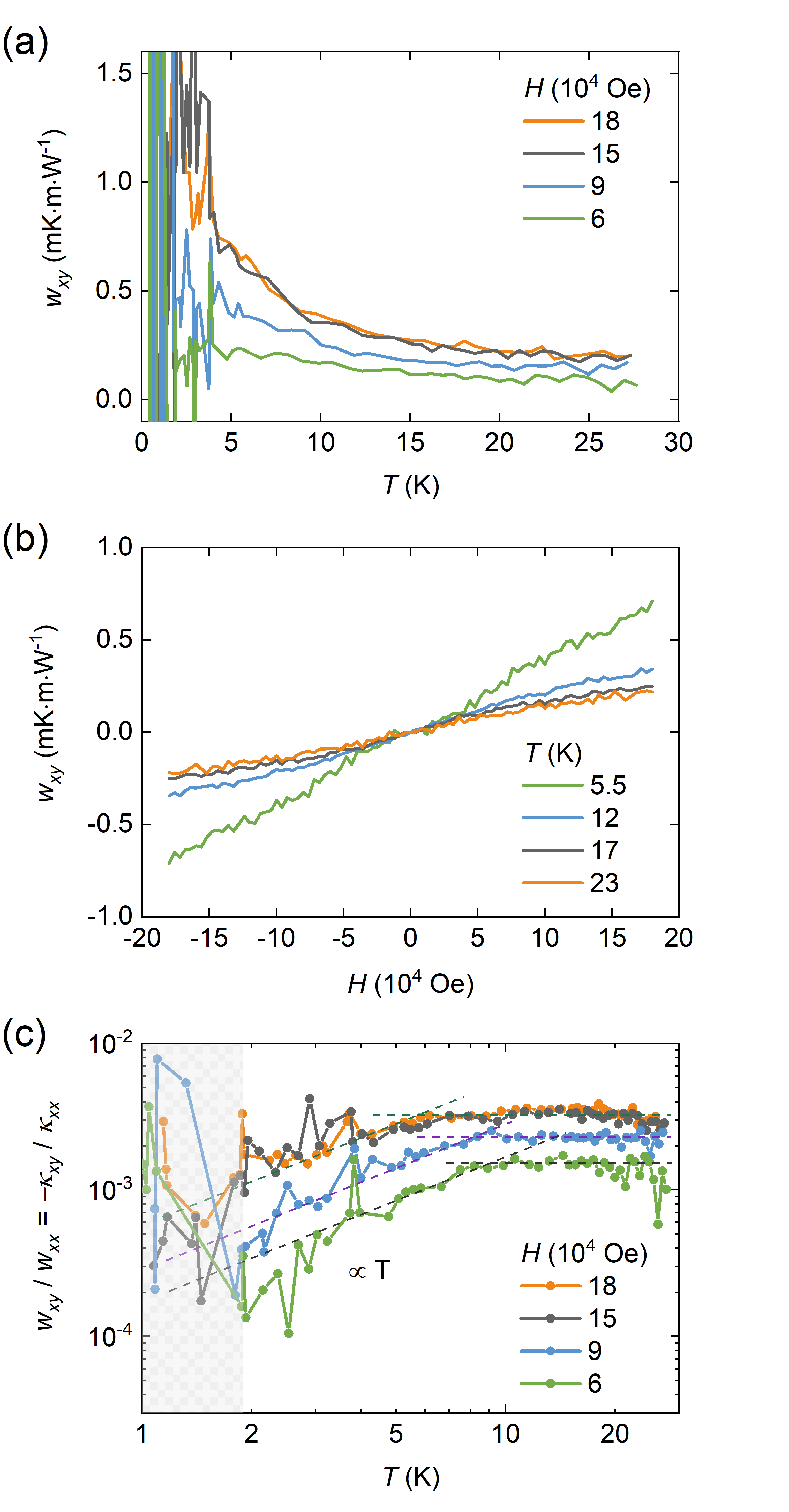}
	\caption{\label{Fig2} Measured thermal Hall resistivity and thermal Hall angle of \LSNOacro. (a)~Temperature dependence of the anti-symmetrized thermal Hall resistance $w_{xy}$ at various magnetic fields following a diverging $T^{-1}$ power-law. (b)~Linear magnetic-field dependence of $w_{xy}$ near $H=0$ at different temperatures. (c)~$T$-linear temperature dependence of the thermal Hall angle at low temperatures. These three phenomena can be found prevalently in other insulators showing phonon THE.}
\end{figure}

In stark contrast, we demonstrate that in \LSNOacro\ the directly measured thermal resistivity tensor differentiates a transverse phonon THE from the longitudinal magneto-thermal effect. Figure~\ref{Fig2}(a) shows the thermal Hall resistivity $w_{xy}$ versus temperature, illustrating an absence of any peak feature as compared to $\kappa_{xy}$ (\textit{cf.} Fig.~\ref{Fig1}(a)). We emphasize that $w_{xy}(T)$ remains an insulating-like diverging function $\propto T^{-1}$, before a diminishing measurement sensitivity of $w_{xy}$ below $\sim 4$ K. At higher temperatures, $w_{xy}/H$ appears to level off at $\sim 10^{-5}\ \text{K}{\cdot}\text{m}{\cdot}\text{W}^{-1}{\cdot}\text{T}^{-1}$ \cite{supp}, similar to the density-wave-ordered trilayer nickelate \cite{xiang_density-wave_2026} and consistent with the previous observation of a unifying magnitude of $w_{xy}$ \cite{guo_interaction_2026}. The apparent difference between $w_{xy}$ and $\kappa_{xy}$ suggests that $w_{xy}$ probes the transverse response related to phonon THE and is less sensitive to longitudinal thermal transport, as predicted in Refs.~\cite{guo_resonant_2022, mangeolle_phonon_2022}.

Furthermore, in Fig.~\ref{Fig2}(b), we plot the magnetic-field dependence of $w_{xy}$ at different temperatures, which is predominantly $H$-linear especially near $H=0$. The thermal Hall coefficient $dw_{xy}/dH$ as $H\to0$ was found to exhibit a similar $T^{-1}$ divergence \cite{supp} as in $w_{xy}(T)$ at 18 T. Meanwhile, the thermal Hall angle as shown in Fig.~\ref{Fig2}(c), defined as approximately $|\kappa_{xy}|/\kappa_{xx}$, or equivalently $w_{xy}/w_{xx}$, follows a $T$-linear power-law at low temperatures $\lesssim 5$ K and then enters a peak above $\sim 10$ K. The magnitude of the thermal Hall angle peak in \LSNOacro\ is comparable to almost any materials that exhibit phonon THE \cite{boulanger_thermal_2020, guo_interaction_2026}. The $T$-linear thermal Hall angle directly results from the ratio between the power-law temperature dependence of $\kappa_{xy}/\kappa_{xx}\propto T^3/T^2\sim T$ at low temperatures.	Overall, the thermal Hall resistivity $w_{xy}$ of \LSNOacro\ shows a monotonic diverging temperature dependence, a linear magnetic-field dependence near $H=0$, and a $T$-linear thermal Hall angle.

\begin{figure}[!tb]
	\centering
	\includegraphics[width=\columnwidth]{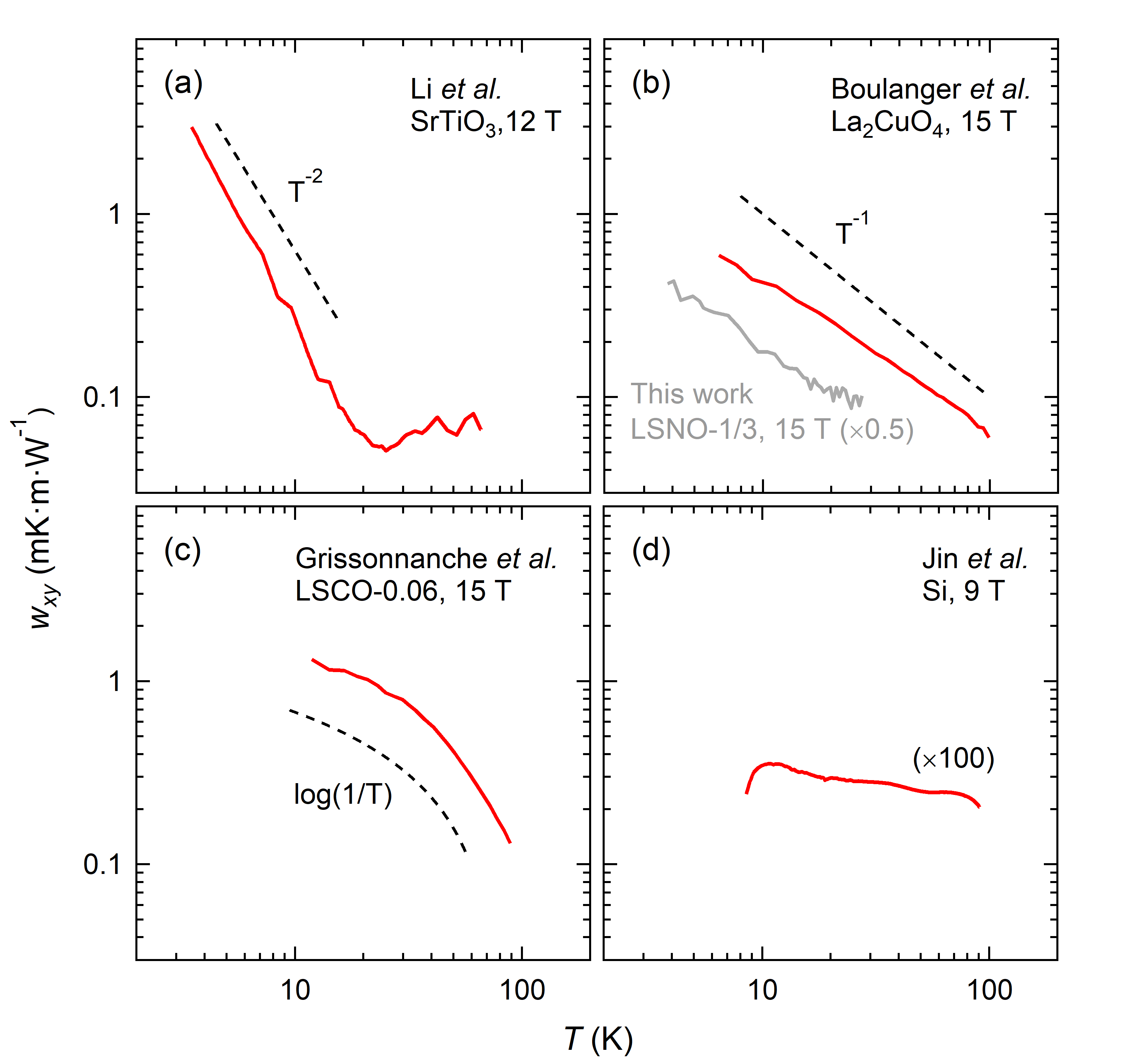}
	\caption{\label{Fig3} Insulating-like temperature dependence of thermal Hall resistivity $w_{xy}=-\kappa_{xy}/\kappa_{xx}^2$ at low temperatures of (a) SrTiO$_3$ from Ref.~\cite{li_phonon_2020} showing a $T^{-2}$ divergence, (b) La$_2$CuO$_4$ from Ref.~\cite{boulanger_thermal_2020} showing a $T^{-1}$ divergence similar to \LSNOacro\ in this work, (c) LSCO-0.06 from Ref.~\cite{grissonnanche_giant_2019} showing a logarithmic divergence, and (d) Silicon from Ref.~\cite{jin_discovery_2025} showing a weak negative temperature dependence. The $w_{xy}$ of silicon was calculated using $\kappa_{xx}(H=0)$ due to data availability; data was multiplied by 100 to facilitate the comparison with other materials. All dashed lines are guides to the eye, indicating the respective divergence laws.}
\end{figure}

We propose as our central result that these three phenomena of the themal Hall resistivity $w_{xy}$ form a set of unifying descriptions of phonon THE in a wide range of insulating materials, and we examine their commonalities and subtle differences in the following.

\textit{Insulating-like $w_{xy}(T)$:} in Fig.~\ref{Fig3}, we plot the thermal Hall resistivity as extracted and calculated from Refs.~\cite{grissonnanche_giant_2019, li_phonon_2020, boulanger_thermal_2020, jin_discovery_2025}, in comparison with our measured $w_{xy}(T)$ of \LSNOacro. A diverse collection of insulating-like $w_{xy}(T)$ is shown to persist over a broad temperature range. In the non-magnetic band-insulator STO \cite{collignon_metallicity_2019, li_phonon_2020} (Fig.~\ref{Fig3}(a)), the low-temperature $T^{-2}$ divergence is stronger than the $T^{-1}$ divergence of \LSNOacro, as a direct consequence of the combination of observed power-laws in $w_{xy}(T)=-\kappa_{xy}(T)/\kappa_{xx}^2(T)\propto T^4/(T^3)^2=T^{-2}$ for STO or $\propto T^3/(T^2)^2=T^{-1}$ for \LSNOacro. In the cuprate parental compound La$_2$CuO$_4$ (LCO) \cite{boulanger_thermal_2020} (Fig.~\ref{Fig3}(b)), a 3d antiferromagnetic (AFM) Mott insulator which likely includes some unavoidable oxygen doping \cite{cheong_properties_1989}, a $T^{-1}$ divergence over the entire temperature range strongly resembles the $w_{xy}$ divergence in our measured \LSNOacro. Whereas \LSNOacro\ has only quasi-1d AFM correlations \cite{freeman_low-energy_2011}, it shares a low-temperature variable-range-hopping-dominated electronic state with LCO \cite{cheong_properties_1989, matsushita_electrical_1990}. The trend of a weakening divergence of $w_{xy}$ extends to the underdoped cuprate La$_{1.94}$Sr$_{0.06}$CuO$_4$ (LSCO-0.06) \cite{grissonnanche_giant_2019} (Fig.~\ref{Fig3}(c)), which hosts a pseudogap regime, superconductivity below $\approx 5$ K in zero magnetic field, and a high-field insulating ground state. We can fit a logarithmic divergence $w_{xy}\propto\log(1/T)$ over a broad temperature range \cite{supp}. A comparable logarithmic divergence has been scrutinized in the electrical resistivity of the high-field insulating ground state of LSCO-0.06 \cite{ando_logarithmic_1995}. Lastly, in the semiconducting silicon \cite{jin_discovery_2025} (Figure~\ref{Fig3}(d)), except the rapid downturn below $\approx 10$ K \footnote{The low-temperature downturn in silicon may result from calculating the thermal Hall resistivity using $\kappa_{xy}(9\text{T})$ and $\kappa_{xx}(0\text{T})$, as $\kappa_{xx}(9\text{T})$ was not provided.}, a weak negative temperature dependence in the thermal Hall resistivity can be seen over most of the temperature range, possibly related to extrinsic impurity scattering, which was known to strongly affect its electrical and thermal conductivity \cite{shklovskii_electronic_1984}.

\textit{Linear $w_{xy}(H)$ near $H=0$:} a low-field linear magnetic-field dependence of the Hall resistivity is a defining feature of the conventional Hall effect. In previous THE studies, linear $w_{xy}(H)$ was first noted (although plotted as the transverse temperature difference $\Delta T_y$) in the paramagnetic insulator Tb$_3$Ga$_5$O$_{12}$, the earliest proposed manifestation of the phonon Hall effect \cite{strohm_phenomenological_2005}. Similarly, in STO, the magnetic-field dependence of $\Delta T_y\propto w_{xy}$ is remarkably linear up to at least $H=12$ T \cite{li_phonon_2020} as a canonical signature of phonon THE. 

\textit{$T$-linear thermal Hall angle:} a low-temperature $T$-linear thermal Hall angle in STO results from the separately identified power-law temperature dependence of $\kappa_{xy}\propto T^4$ and $\kappa_{xx}\propto T^3$ \cite{li_phonon_2020}, whereas in semiconducting silicon and germanium, the thermal Hall angle versus temperature as extracted from Ref.~\cite{jin_discovery_2025} also features a linear trend below the intermediate-temperature phononic peak \cite{supp}.

Having presented $w_{xy}$ as a unifying description of phonon THE, we discuss potential physical implications of the observed correlation between the insulating-like $w_{xy}(T)$ and material's localized electronic state. Theoretical models relating phonon THE to localized charge degrees of freedom typically involve an asymmetric scattering between phonons and extrinsic charge defects or impurities \cite{mangeolle_phonon_2022, oh_phonon_2025}. Large phonon THE of $\kappa_{xy}/\kappa_{xx}\sim 10^{-3}$ has been predicted, involving resonant skew or side-jump scattering \cite{guo_resonant_2022, sun_large_2022} and phonon skew-scattering with charged defects that are subject to the Lorentz force \cite{flebus_charged_2022}. Recently, Refs.~\cite{guo_interaction_2026, ling_phonon_2026} showed that THE can arise from the magnetic-field-induced geometric phase of an unbalanced charge-nuclei distribution, predicting a unifying magnitude of $w_{xy}$ for various insulators and semiconductors. Extending beyond microscopic pictures, a macroscopic coupling between a charged ion and a charge-neutral harmonic oscillator (phonon) bath, based on the Caldeira-Leggett model, can facilitate the transfer of finite angular momentum from charged ions to the phonon bath in thermal equilibrium \cite{matevosyan_lasting_2023}. These models exemplify the diffusion of angular momentum from charge degrees of freedom to heat-carrying lattice vibrations in insulators, potentially yielding unifying phenomena of phonon THE in various material systems.

If phonon motion were to acquire chirality from its coupling to localized charges, one would expect that the phonon Hall transport inherits certain features of the underlying electronic state. The fact that the $w_{xy}$ power-law in \LSNOacro\ is weaker than that in STO may be related to a 2d phonon specific heat $c\propto T^2$ in \LSNOacro\ (\textit{vs.} $c\propto T^3$ for 3d phonons), manifested as a carrier-related concave hump in $c/T$ vs. $T^2$ at low temperatures \cite{matsushita_electrical_1990}. Thus, the phonon thermal (Hall) transport appears to be sensitive to the quasi-2d charge distribution. As for \LSCOacro-0.06, since the high-field logarithmic resistivity divergence could be linked to similar behaviors in granular systems \cite{boebinger_insulator--metal_1996, steiner_possible_2005}, its potential origin could affect the thermal Hall transport alike. Similar arguments can be applied to semiconducting silicon where impurity scattering closely controls both charge and phonon Hall transport. It would be intriguing to investigate how the unifying description of the phonon thermal Hall resistivity influences the modeling of magneto-thermal transport in various insulating systems, ranging from crystalline solids to soft condensed matters.

Lastly, we examine possible magnetic contributions to THE in \LSNOacro\ which hosts disordered quasi-1d AFM stripes \cite{freeman_magnetization_2006, freeman_low-energy_2011}. Whereas magnon-dominated THE \cite{katsura_theory_2010} often involves robust long-range magnetic order \cite{onose_observation_2010, ideue_giant_2017, zhang_thermal_2019, zhang_anomalous_2021}, phonons were shown to dominate the THE in LCO \cite{grissonnanche_chiral_2020}, a 3d AFM. While the present data cannot definitely rule out the phonon-magnon contribution \cite{sugii_thermal_2017, zhang_thermal_2019, zhang_anomalous_2021}, we can contrive two scenarios to understand comparable THE in both magnetic and non-magnetic insulators: (a) the phonon-dominated THE is amplified by the background magnetic ordering \cite{boulanger_thermal_2026, xiang_density-wave_2026} or a magneto-electric coupling \cite{jiang_dielectric_2020} or (b) phonon-charge and phonon-magnon contributions yield similar THE phenomena of different origins. Future studies are warranted to distinguish these scenarios by tuning phonon-charge coupling while controlling magnetic influence.

In summary, we studied the phonon THE in doped Mott insulator \LSNOacro\ with comparisons across various insulators. The thermal Hall resistivity $w_{xy}$ was shown to differentiate between the transverse phonon THE and longitudinal contributions. Three characteristic phenomena in the temperature and magnetic-field dependencies of $w_{xy}$ were experimentally demonstrated as a unifying description of phonon THE in a diverse range of insulators and semiconductors. Our finding was further corroborated by an apparent correlation between the evolution of $w_{xy}(T)$ and the corresponding material's electronic state from which the phonon Hall transport inherits certain phenomenological features.

\begin{acknowledgments}
\textit{Acknowledgements---} We thank Kamran Behnia, Jinlyu Cao and Hao Sun for useful discussions. The work in Singapore was supported by the Singapore Ministry of Education (MOE) Academic Research Fund Tier 3 grant MOE-MOET32023-0003 and Tier 2 grant MOE-T2EP50223-0019. A portion of this work was performed at the National High Magnetic Field Laboratory, which is supported by the National Science Foundation Cooperative Agreement No. DMR-2128556 and the State of Florida, and the U. S. Department of Energy.
\end{acknowledgments}

\bibliography{THE_LSNO}

\begin{thebibliography}{24}%
	\makeatletter
	\providecommand \@ifxundefined [1]{%
		\@ifx{#1\undefined}
	}%
	\providecommand \@ifnum [1]{%
		\ifnum #1\expandafter \@firstoftwo
		\else \expandafter \@secondoftwo
		\fi
	}%
	\providecommand \@ifx [1]{%
		\ifx #1\expandafter \@firstoftwo
		\else \expandafter \@secondoftwo
		\fi
	}%
	\providecommand \natexlab [1]{#1}%
	\providecommand \enquote  [1]{``#1''}%
	\providecommand \bibnamefont  [1]{#1}%
	\providecommand \bibfnamefont [1]{#1}%
	\providecommand \citenamefont [1]{#1}%
	\providecommand \href@noop [0]{\@secondoftwo}%
	\providecommand \href [0]{\begingroup \@sanitize@url \@href}%
	\providecommand \@href[1]{\@@startlink{#1}\@@href}%
	\providecommand \@@href[1]{\endgroup#1\@@endlink}%
	\providecommand \@sanitize@url [0]{\catcode `\\12\catcode `\$12\catcode
		`\&12\catcode `\#12\catcode `\^12\catcode `\_12\catcode `\%12\relax}%
	\providecommand \@@startlink[1]{}%
	\providecommand \@@endlink[0]{}%
	\providecommand \url  [0]{\begingroup\@sanitize@url \@url }%
	\providecommand \@url [1]{\endgroup\@href {#1}{\urlprefix }}%
	\providecommand \urlprefix  [0]{URL }%
	\providecommand \Eprint [0]{\href }%
	\providecommand \doibase [0]{https://doi.org/}%
	\providecommand \selectlanguage [0]{\@gobble}%
	\providecommand \bibinfo  [0]{\@secondoftwo}%
	\providecommand \bibfield  [0]{\@secondoftwo}%
	\providecommand \translation [1]{[#1]}%
	\providecommand \BibitemOpen [0]{}%
	\providecommand \bibitemStop [0]{}%
	\providecommand \bibitemNoStop [0]{.\EOS\space}%
	\providecommand \EOS [0]{\spacefactor3000\relax}%
	\providecommand \BibitemShut  [1]{\csname bibitem#1\endcsname}%
	\let\auto@bib@innerbib\@empty
	\bibitem [{\citenamefont {Klingeler}\ \emph {et~al.}(2005)\citenamefont
		{Klingeler}, \citenamefont {B\"uchner}, \citenamefont {Cheong},\ and\
		\citenamefont {H\"ucker}}]{Klingeler2005}%
	\BibitemOpen
	\bibfield  {author} {\bibinfo {author} {\bibfnamefont {R.}~\bibnamefont
			{Klingeler}}, \bibinfo {author} {\bibfnamefont {B.}~\bibnamefont
			{B\"uchner}}, \bibinfo {author} {\bibfnamefont {S.-W.}\ \bibnamefont
			{Cheong}},\ and\ \bibinfo {author} {\bibfnamefont {M.}~\bibnamefont
			{H\"ucker}},\ }\bibfield  {title} {\bibinfo {title} {Weak ferromagnetic spin
			and charge stripe order in
			$\mathrm{La}_{5/3}\mathrm{Sr}_{1/3}\mathrm{Ni}\mathrm{O}_4$},\ }\href
	{https://doi.org/10.1103/PhysRevB.72.104424} {\bibfield  {journal} {\bibinfo
			{journal} {Phys. Rev. B}\ }\textbf {\bibinfo {volume} {72}},\ \bibinfo
		{pages} {104424} (\bibinfo {year} {2005})}\BibitemShut {NoStop}%
	\bibitem [{\citenamefont {Freeman}\ \emph {et~al.}(2006)\citenamefont
		{Freeman}, \citenamefont {Boothroyd}, \citenamefont {Prabhakaran},\ and\
		\citenamefont {Lorenzana}}]{55_Freeman2006}%
	\BibitemOpen
	\bibfield  {author} {\bibinfo {author} {\bibfnamefont {P.~G.}\ \bibnamefont
			{Freeman}}, \bibinfo {author} {\bibfnamefont {A.~T.}\ \bibnamefont
			{Boothroyd}}, \bibinfo {author} {\bibfnamefont {D.}~\bibnamefont
			{Prabhakaran}},\ and\ \bibinfo {author} {\bibfnamefont {J.}~\bibnamefont
			{Lorenzana}},\ }\bibfield  {title} {\bibinfo {title} {Magnetization of
			$\mathrm{La}_{2-x}\mathrm{Sr}_{x}\mathrm{Ni}\mathrm{O}_{4+\delta}$
			$(0\ensuremath{\leqslant}x\ensuremath{\leqslant}0.5)$: Spin-glass and memory
			effects},\ }\href {https://doi.org/10.1103/PhysRevB.73.014434} {\bibfield
		{journal} {\bibinfo  {journal} {Phys. Rev. B}\ }\textbf {\bibinfo {volume}
			{73}},\ \bibinfo {pages} {014434} (\bibinfo {year} {2006})}\BibitemShut
	{NoStop}%
	\bibitem [{\citenamefont {H\"ucker}\ \emph {et~al.}(2004)\citenamefont
		{H\"ucker}, \citenamefont {Chung}, \citenamefont {Chand}, \citenamefont
		{Vogt}, \citenamefont {Tranquada},\ and\ \citenamefont
		{Buttrey}}]{Hucker2004}%
	\BibitemOpen
	\bibfield  {author} {\bibinfo {author} {\bibfnamefont {M.}~\bibnamefont
			{H\"ucker}}, \bibinfo {author} {\bibfnamefont {K.}~\bibnamefont {Chung}},
		\bibinfo {author} {\bibfnamefont {M.}~\bibnamefont {Chand}}, \bibinfo
		{author} {\bibfnamefont {T.}~\bibnamefont {Vogt}}, \bibinfo {author}
		{\bibfnamefont {J.~M.}\ \bibnamefont {Tranquada}},\ and\ \bibinfo {author}
		{\bibfnamefont {D.~J.}\ \bibnamefont {Buttrey}},\ }\bibfield  {title}
	{\bibinfo {title} {Oxygen and strontium codoping of $\mathrm{La}_2
			\mathrm{Ni}\mathrm{O}_4$: Room-temperature phase diagrams},\ }\href
	{https://doi.org/10.1103/PhysRevB.70.064105} {\bibfield  {journal} {\bibinfo
			{journal} {Phys. Rev. B}\ }\textbf {\bibinfo {volume} {70}},\ \bibinfo
		{pages} {064105} (\bibinfo {year} {2004})}\BibitemShut {NoStop}%
	\bibitem [{\citenamefont {Park}\ \emph {et~al.}(2005)\citenamefont {Park},
		\citenamefont {Nussinov}, \citenamefont {Hazzard}, \citenamefont {Sidorov},
		\citenamefont {Balatsky}, \citenamefont {Sarrao}, \citenamefont {Cheong},
		\citenamefont {Hundley}, \citenamefont {Lee}, \citenamefont {Jia},\ and\
		\citenamefont {Thompson}}]{53_Park2005}%
	\BibitemOpen
	\bibfield  {author} {\bibinfo {author} {\bibfnamefont {T.}~\bibnamefont
			{Park}}, \bibinfo {author} {\bibfnamefont {Z.}~\bibnamefont {Nussinov}},
		\bibinfo {author} {\bibfnamefont {K.~R.~A.}\ \bibnamefont {Hazzard}},
		\bibinfo {author} {\bibfnamefont {V.~A.}\ \bibnamefont {Sidorov}}, \bibinfo
		{author} {\bibfnamefont {A.~V.}\ \bibnamefont {Balatsky}}, \bibinfo {author}
		{\bibfnamefont {J.~L.}\ \bibnamefont {Sarrao}}, \bibinfo {author}
		{\bibfnamefont {S.-W.}\ \bibnamefont {Cheong}}, \bibinfo {author}
		{\bibfnamefont {M.~F.}\ \bibnamefont {Hundley}}, \bibinfo {author}
		{\bibfnamefont {J.-S.}\ \bibnamefont {Lee}}, \bibinfo {author} {\bibfnamefont
			{Q.~X.}\ \bibnamefont {Jia}},\ and\ \bibinfo {author} {\bibfnamefont {J.~D.}\
			\bibnamefont {Thompson}},\ }\bibfield  {title} {\bibinfo {title} {Novel
			dielectric anomaly in the hole-doped
			$\mathrm{La}_{2}\mathrm{Cu}_{1-x}\mathrm{Li}_x \mathrm{O}_4$ and
			$\mathrm{La}_{2-x}\mathrm{Sr}_{x}\mathrm{Ni}\mathrm{O}_4$ insulators:
			Signature of an electronic glassy state},\ }\href
	{https://doi.org/10.1103/PhysRevLett.94.017002} {\bibfield  {journal}
		{\bibinfo  {journal} {Phys. Rev. Lett.}\ }\textbf {\bibinfo {volume} {94}},\
		\bibinfo {pages} {017002} (\bibinfo {year} {2005})}\BibitemShut {NoStop}%
	\bibitem [{\citenamefont {Spencer}\ \emph {et~al.}(2005)\citenamefont
		{Spencer}, \citenamefont {Ghazi}, \citenamefont {Wilkins}, \citenamefont
		{Hatton}, \citenamefont {Brown}, \citenamefont {Prabhakaran},\ and\
		\citenamefont {Boothroyd}}]{54_Spencer2005}%
	\BibitemOpen
	\bibfield  {author} {\bibinfo {author} {\bibfnamefont {P.~D.}\ \bibnamefont
			{Spencer}}, \bibinfo {author} {\bibfnamefont {M.~E.}\ \bibnamefont {Ghazi}},
		\bibinfo {author} {\bibfnamefont {S.~B.}\ \bibnamefont {Wilkins}}, \bibinfo
		{author} {\bibfnamefont {P.~D.}\ \bibnamefont {Hatton}}, \bibinfo {author}
		{\bibfnamefont {S.~D.}\ \bibnamefont {Brown}}, \bibinfo {author}
		{\bibfnamefont {D.}~\bibnamefont {Prabhakaran}},\ and\ \bibinfo {author}
		{\bibfnamefont {A.~T.}\ \bibnamefont {Boothroyd}},\ }\bibfield  {title}
	{\bibinfo {title} {Charge stripe glasses in
			$\mathrm{La}_{2-x}\mathrm{Sr}_{X}\mathrm{Ni}\mathrm{O}_4$ for $0.20 < x <
			0.25$},\ }\href {https://doi.org/10.1140/epjb/e2005-00231-3} {\bibfield
		{journal} {\bibinfo  {journal} {Eur. Phys. J. B}\ }\textbf {\bibinfo {volume}
			{46}},\ \bibinfo {pages} {27} (\bibinfo {year} {2005})}\BibitemShut {NoStop}%
	\bibitem [{\citenamefont {Filippi}\ \emph {et~al.}(2009)\citenamefont
		{Filippi}, \citenamefont {Kundys}, \citenamefont {Agrestini}, \citenamefont
		{Prellier}, \citenamefont {Oyanagi},\ and\ \citenamefont
		{Saini}}]{56_Filippi2009}%
	\BibitemOpen
	\bibfield  {author} {\bibinfo {author} {\bibfnamefont {M.}~\bibnamefont
			{Filippi}}, \bibinfo {author} {\bibfnamefont {B.}~\bibnamefont {Kundys}},
		\bibinfo {author} {\bibfnamefont {S.}~\bibnamefont {Agrestini}}, \bibinfo
		{author} {\bibfnamefont {W.}~\bibnamefont {Prellier}}, \bibinfo {author}
		{\bibfnamefont {H.}~\bibnamefont {Oyanagi}},\ and\ \bibinfo {author}
		{\bibfnamefont {N.~L.}\ \bibnamefont {Saini}},\ }\bibfield  {title} {\bibinfo
		{title} {Charge order, dielectric response, and local structure of
			$\mathrm{La}_{5/3}\mathrm{Sr}_{1/3}\mathrm{Ni}\mathrm{O}_4$ system},\ }\href
	{https://doi.org/10.1063/1.3260222} {\bibfield  {journal} {\bibinfo
			{journal} {J. Appl. Phys.}\ }\textbf {\bibinfo {volume} {106}},\ \bibinfo
		{pages} {104116} (\bibinfo {year} {2009})}\BibitemShut {NoStop}%
	\bibitem [{\citenamefont {Chen}\ \emph {et~al.}(1993)\citenamefont {Chen},
		\citenamefont {Cheong},\ and\ \citenamefont {Cooper}}]{47_CHChen1993}%
	\BibitemOpen
	\bibfield  {author} {\bibinfo {author} {\bibfnamefont {C.~H.}\ \bibnamefont
			{Chen}}, \bibinfo {author} {\bibfnamefont {S.-W.}\ \bibnamefont {Cheong}},\
		and\ \bibinfo {author} {\bibfnamefont {A.~S.}\ \bibnamefont {Cooper}},\
	}\bibfield  {title} {\bibinfo {title} {Charge modulations in
			$\mathrm{La}_{2-x}\mathrm{Sr}_{x}\mathrm{Ni}\mathrm{O}_{4+y}$: Ordering of
			polarons},\ }\href {https://doi.org/10.1103/PhysRevLett.71.2461} {\bibfield
		{journal} {\bibinfo  {journal} {Phys. Rev. Lett.}\ }\textbf {\bibinfo
			{volume} {71}},\ \bibinfo {pages} {2461} (\bibinfo {year}
		{1993})}\BibitemShut {NoStop}%
	\bibitem [{\citenamefont {McQueeney}\ \emph {et~al.}(1999)\citenamefont
		{McQueeney}, \citenamefont {Sarrao},\ and\ \citenamefont
		{Osborn}}]{57_McQueeney1999}%
	\BibitemOpen
	\bibfield  {author} {\bibinfo {author} {\bibfnamefont {R.~J.}\ \bibnamefont
			{McQueeney}}, \bibinfo {author} {\bibfnamefont {J.~L.}\ \bibnamefont
			{Sarrao}},\ and\ \bibinfo {author} {\bibfnamefont {R.}~\bibnamefont
			{Osborn}},\ }\bibfield  {title} {\bibinfo {title} {Phonon densities of states
			of $\mathrm{La}_{2-x}\mathrm{Sr}_{x}\mathrm{Ni}\mathrm{O}_{4}$ evidence for
			strong electron-lattice coupling},\ }\href
	{https://doi.org/10.1103/PhysRevB.60.80} {\bibfield  {journal} {\bibinfo
			{journal} {Phys. Rev. B}\ }\textbf {\bibinfo {volume} {60}},\ \bibinfo
		{pages} {80} (\bibinfo {year} {1999})}\BibitemShut {NoStop}%
	\bibitem [{\citenamefont {Merritt}\ \emph {et~al.}(2019)\citenamefont
		{Merritt}, \citenamefont {Reznik}, \citenamefont {Garlea}, \citenamefont
		{Gu},\ and\ \citenamefont {Tranquada}}]{Merritt2019}%
	\BibitemOpen
	\bibfield  {author} {\bibinfo {author} {\bibfnamefont {A.~M.}\ \bibnamefont
			{Merritt}}, \bibinfo {author} {\bibfnamefont {D.}~\bibnamefont {Reznik}},
		\bibinfo {author} {\bibfnamefont {V.~O.}\ \bibnamefont {Garlea}}, \bibinfo
		{author} {\bibfnamefont {G.~D.}\ \bibnamefont {Gu}},\ and\ \bibinfo {author}
		{\bibfnamefont {J.~M.}\ \bibnamefont {Tranquada}},\ }\bibfield  {title}
	{\bibinfo {title} {Nature and impact of stripe freezing in
			$\mathrm{La}_{1.67}\mathrm{Sr}_{0.33}\mathrm{Ni}\mathrm{O}_{4}$},\ }\href
	{https://doi.org/10.1103/PhysRevB.100.195122} {\bibfield  {journal} {\bibinfo
			{journal} {Phys. Rev. B}\ }\textbf {\bibinfo {volume} {100}},\ \bibinfo
		{pages} {195122} (\bibinfo {year} {2019})}\BibitemShut {NoStop}%
	\bibitem [{\citenamefont {Freeman}\ \emph {et~al.}(2011)\citenamefont
		{Freeman}, \citenamefont {Prabhakaran}, \citenamefont {Nakajima},
		\citenamefont {Stunault}, \citenamefont {Enderle}, \citenamefont
		{Niedermayer}, \citenamefont {Frost}, \citenamefont {Yamada},\ and\
		\citenamefont {Boothroyd}}]{58_Freeman2011}%
	\BibitemOpen
	\bibfield  {author} {\bibinfo {author} {\bibfnamefont {P.~G.}\ \bibnamefont
			{Freeman}}, \bibinfo {author} {\bibfnamefont {D.}~\bibnamefont
			{Prabhakaran}}, \bibinfo {author} {\bibfnamefont {K.}~\bibnamefont
			{Nakajima}}, \bibinfo {author} {\bibfnamefont {A.}~\bibnamefont {Stunault}},
		\bibinfo {author} {\bibfnamefont {M.}~\bibnamefont {Enderle}}, \bibinfo
		{author} {\bibfnamefont {C.}~\bibnamefont {Niedermayer}}, \bibinfo {author}
		{\bibfnamefont {C.~D.}\ \bibnamefont {Frost}}, \bibinfo {author}
		{\bibfnamefont {K.}~\bibnamefont {Yamada}},\ and\ \bibinfo {author}
		{\bibfnamefont {A.~T.}\ \bibnamefont {Boothroyd}},\ }\bibfield  {title}
	{\bibinfo {title} {Low-energy quasi-one-dimensional spin dynamics in
			charge-ordered $\mathrm{La}_{2-x}\mathrm{Sr}_{x}\mathrm{Ni}\mathrm{O}_{4}$},\
	}\href {https://doi.org/10.1103/PhysRevB.83.094414} {\bibfield  {journal}
		{\bibinfo  {journal} {Phys. Rev. B}\ }\textbf {\bibinfo {volume} {83}},\
		\bibinfo {pages} {094414} (\bibinfo {year} {2011})}\BibitemShut {NoStop}%
	\bibitem [{\citenamefont {Zachar}\ \emph {et~al.}(1998)\citenamefont {Zachar},
		\citenamefont {Kivelson},\ and\ \citenamefont {Emery}}]{50_Zachar1998}%
	\BibitemOpen
	\bibfield  {author} {\bibinfo {author} {\bibfnamefont {O.}~\bibnamefont
			{Zachar}}, \bibinfo {author} {\bibfnamefont {S.~A.}\ \bibnamefont
			{Kivelson}},\ and\ \bibinfo {author} {\bibfnamefont {V.~J.}\ \bibnamefont
			{Emery}},\ }\bibfield  {title} {\bibinfo {title} {Landau theory of stripe
			phases in cuprates and nickelates},\ }\href
	{https://doi.org/10.1103/PhysRevB.57.1422} {\bibfield  {journal} {\bibinfo
			{journal} {Phys. Rev. B}\ }\textbf {\bibinfo {volume} {57}},\ \bibinfo
		{pages} {1422} (\bibinfo {year} {1998})}\BibitemShut {NoStop}%
	\bibitem [{\citenamefont {Matsushita}\ \emph {et~al.}(1990)\citenamefont
		{Matsushita}, \citenamefont {Matsumoto}, \citenamefont {Takayanagi},\ and\
		\citenamefont {M{\=o}ri}}]{MATSUSHITA1990}%
	\BibitemOpen
	\bibfield  {author} {\bibinfo {author} {\bibfnamefont {A.}~\bibnamefont
			{Matsushita}}, \bibinfo {author} {\bibfnamefont {T.}~\bibnamefont
			{Matsumoto}}, \bibinfo {author} {\bibfnamefont {S.}~\bibnamefont
			{Takayanagi}},\ and\ \bibinfo {author} {\bibfnamefont {N.}~\bibnamefont
			{M{\=o}ri}},\ }\bibfield  {title} {\bibinfo {title} {{Electrical resistivity
				and specific heat of La$_{2- X}$Sr$_X$NiO$_{4+\delta}$}},\ }\href@noop {}
	{\bibfield  {journal} {\bibinfo  {journal} {Phys. B: Condens. Matter.}\
		}\textbf {\bibinfo {volume} {165}},\ \bibinfo {pages} {1351} (\bibinfo {year}
		{1990})}\BibitemShut {NoStop}%
	\bibitem [{\citenamefont {Hess}\ \emph {et~al.}(1999)\citenamefont {Hess},
		\citenamefont {B\"uchner}, \citenamefont {H\"ucker}, \citenamefont {Gross},\
		and\ \citenamefont {Cheong}}]{Hess1999}%
	\BibitemOpen
	\bibfield  {author} {\bibinfo {author} {\bibfnamefont {C.}~\bibnamefont
			{Hess}}, \bibinfo {author} {\bibfnamefont {B.}~\bibnamefont {B\"uchner}},
		\bibinfo {author} {\bibfnamefont {M.}~\bibnamefont {H\"ucker}}, \bibinfo
		{author} {\bibfnamefont {R.}~\bibnamefont {Gross}},\ and\ \bibinfo {author}
		{\bibfnamefont {S.-W.}\ \bibnamefont {Cheong}},\ }\bibfield  {title}
	{\bibinfo {title} {Phonon thermal conductivity and stripe correlations in
			$\mathrm{La}_{2-x}\mathrm{Sr}_{x}\mathrm{Ni}\mathrm{O}_{4}$ and
			$\mathrm{Sr}_{1.5}\mathrm{La}_{0.5}\mathrm{Mn}\mathrm{O}_{4}$},\ }\href
	{https://doi.org/10.1103/PhysRevB.59.R10397} {\bibfield  {journal} {\bibinfo
			{journal} {Phys. Rev. B}\ }\textbf {\bibinfo {volume} {59}},\ \bibinfo
		{pages} {R10397} (\bibinfo {year} {1999})}\BibitemShut {NoStop}%
	\bibitem [{\citenamefont {Yoshizawa}\ \emph {et~al.}(2000)\citenamefont
		{Yoshizawa}, \citenamefont {Kakeshita}, \citenamefont {Kajimoto},
		\citenamefont {Tanabe}, \citenamefont {Katsufuji},\ and\ \citenamefont
		{Tokura}}]{51_Yoshizawa2000}%
	\BibitemOpen
	\bibfield  {author} {\bibinfo {author} {\bibfnamefont {H.}~\bibnamefont
			{Yoshizawa}}, \bibinfo {author} {\bibfnamefont {T.}~\bibnamefont
			{Kakeshita}}, \bibinfo {author} {\bibfnamefont {R.}~\bibnamefont {Kajimoto}},
		\bibinfo {author} {\bibfnamefont {T.}~\bibnamefont {Tanabe}}, \bibinfo
		{author} {\bibfnamefont {T.}~\bibnamefont {Katsufuji}},\ and\ \bibinfo
		{author} {\bibfnamefont {Y.}~\bibnamefont {Tokura}},\ }\bibfield  {title}
	{\bibinfo {title} {{Stripe order at low temperatures in
				${\mathrm{La}}_{2\ensuremath{-}x}{\mathrm{Sr}}_{x}{\mathrm{NiO}}_{4}$ with
				$0.289\ensuremath{\lesssim}x\ensuremath{\lesssim}0.5$}},\ }\href
	{https://doi.org/10.1103/PhysRevB.61.R854} {\bibfield  {journal} {\bibinfo
			{journal} {Phys. Rev. B}\ }\textbf {\bibinfo {volume} {61}},\ \bibinfo
		{pages} {R854(R)} (\bibinfo {year} {2000})}\BibitemShut {NoStop}%
	\bibitem [{\citenamefont {Anissimova}\ \emph {et~al.}(2014)\citenamefont
		{Anissimova}, \citenamefont {Parshall}, \citenamefont {Gu}, \citenamefont
		{Marty}, \citenamefont {Lumsden}, \citenamefont {Chi}, \citenamefont
		{Fernandez-Baca}, \citenamefont {Abernathy}, \citenamefont {Lamago},
		\citenamefont {Tranquada},\ and\ \citenamefont
		{Reznik}}]{Anissimova2014LSNO}%
	\BibitemOpen
	\bibfield  {author} {\bibinfo {author} {\bibfnamefont {S.}~\bibnamefont
			{Anissimova}}, \bibinfo {author} {\bibfnamefont {D.}~\bibnamefont
			{Parshall}}, \bibinfo {author} {\bibfnamefont {G.}~\bibnamefont {Gu}},
		\bibinfo {author} {\bibfnamefont {K.}~\bibnamefont {Marty}}, \bibinfo
		{author} {\bibfnamefont {M.~D.}\ \bibnamefont {Lumsden}}, \bibinfo {author}
		{\bibfnamefont {S.}~\bibnamefont {Chi}}, \bibinfo {author} {\bibfnamefont
			{J.~A.}\ \bibnamefont {Fernandez-Baca}}, \bibinfo {author} {\bibfnamefont
			{D.}~\bibnamefont {Abernathy}}, \bibinfo {author} {\bibfnamefont
			{D.}~\bibnamefont {Lamago}}, \bibinfo {author} {\bibfnamefont {J.~M.}\
			\bibnamefont {Tranquada}},\ and\ \bibinfo {author} {\bibfnamefont
			{D.}~\bibnamefont {Reznik}},\ }\bibfield  {title} {\bibinfo {title} {{Direct
				observation of dynamic charge stripes in La$_{2-x}$Sr$_x$NiO$_4$}},\
	}\href@noop {} {\bibfield  {journal} {\bibinfo  {journal} {Nat. Commun.}\
		}\textbf {\bibinfo {volume} {5}},\ \bibinfo {pages} {3467} (\bibinfo {year}
		{2014})}\BibitemShut {NoStop}%
	\bibitem [{\citenamefont {Lee}\ \emph {et~al.}(2001)\citenamefont {Lee},
		\citenamefont {Cheong}, \citenamefont {Yamada},\ and\ \citenamefont
		{Majkrzak}}]{Lee2001LSNO}%
	\BibitemOpen
	\bibfield  {author} {\bibinfo {author} {\bibfnamefont {S.-H.}\ \bibnamefont
			{Lee}}, \bibinfo {author} {\bibfnamefont {S.-W.}\ \bibnamefont {Cheong}},
		\bibinfo {author} {\bibfnamefont {K.}~\bibnamefont {Yamada}},\ and\ \bibinfo
		{author} {\bibfnamefont {C.}~\bibnamefont {Majkrzak}},\ }\bibfield  {title}
	{\bibinfo {title} {{Charge and canted spin order in La $_{2- x}$Sr$_x$NiO$_4$
				($x= 0.275$ and $\frac{1}{3}$)}},\ }\href@noop {} {\bibfield  {journal}
		{\bibinfo  {journal} {Phys. Rev. B}\ }\textbf {\bibinfo {volume} {63}},\
		\bibinfo {pages} {060405} (\bibinfo {year} {2001})}\BibitemShut {NoStop}%
	\bibitem [{\citenamefont {Ramirez}\ \emph {et~al.}(1996)\citenamefont
		{Ramirez}, \citenamefont {Gammel}, \citenamefont {Cheong}, \citenamefont
		{Bishop},\ and\ \citenamefont {Chandra}}]{Ramirez1996}%
	\BibitemOpen
	\bibfield  {author} {\bibinfo {author} {\bibfnamefont {A.~P.}\ \bibnamefont
			{Ramirez}}, \bibinfo {author} {\bibfnamefont {P.~L.}\ \bibnamefont {Gammel}},
		\bibinfo {author} {\bibfnamefont {S.-W.}\ \bibnamefont {Cheong}}, \bibinfo
		{author} {\bibfnamefont {D.~J.}\ \bibnamefont {Bishop}},\ and\ \bibinfo
		{author} {\bibfnamefont {P.}~\bibnamefont {Chandra}},\ }\bibfield  {title}
	{\bibinfo {title} {Charge modulation in
			$\mathrm{La}_{1.67}\mathrm{Sr}_{0.33}\mathrm{Ni}\mathrm{O}_{4}$: A bulk
			thermodynamic study},\ }\href {https://doi.org/10.1103/PhysRevLett.76.447}
	{\bibfield  {journal} {\bibinfo  {journal} {Phys. Rev. Lett.}\ }\textbf
		{\bibinfo {volume} {76}},\ \bibinfo {pages} {447} (\bibinfo {year}
		{1996})}\BibitemShut {NoStop}%
	\bibitem [{\citenamefont {Jin}\ \emph {et~al.}(2025)\citenamefont {Jin},
		\citenamefont {Zhang}, \citenamefont {Wan}, \citenamefont {Wang},
		\citenamefont {Jiao},\ and\ \citenamefont {Li}}]{Jin2025Silicon}%
	\BibitemOpen
	\bibfield  {author} {\bibinfo {author} {\bibfnamefont {X.}~\bibnamefont
			{Jin}}, \bibinfo {author} {\bibfnamefont {X.}~\bibnamefont {Zhang}}, \bibinfo
		{author} {\bibfnamefont {W.}~\bibnamefont {Wan}}, \bibinfo {author}
		{\bibfnamefont {H.}~\bibnamefont {Wang}}, \bibinfo {author} {\bibfnamefont
			{Y.}~\bibnamefont {Jiao}},\ and\ \bibinfo {author} {\bibfnamefont
			{S.}~\bibnamefont {Li}},\ }\bibfield  {title} {\bibinfo {title} {{Discovery
				of Universal Phonon Thermal Hall Effect in Crystals}},\ }\href@noop {}
	{\bibfield  {journal} {\bibinfo  {journal} {Phys. Rev. Lett.}\ }\textbf
		{\bibinfo {volume} {135}},\ \bibinfo {pages} {196302} (\bibinfo {year}
		{2025})}\BibitemShut {NoStop}%
	\bibitem [{\citenamefont {Li}\ \emph {et~al.}(2020)\citenamefont {Li},
		\citenamefont {Fauqu\'e}, \citenamefont {Zhu},\ and\ \citenamefont
		{Behnia}}]{XiaoKang2020}%
	\BibitemOpen
	\bibfield  {author} {\bibinfo {author} {\bibfnamefont {X.}~\bibnamefont
			{Li}}, \bibinfo {author} {\bibfnamefont {B.}~\bibnamefont {Fauqu\'e}},
		\bibinfo {author} {\bibfnamefont {Z.}~\bibnamefont {Zhu}},\ and\ \bibinfo
		{author} {\bibfnamefont {K.}~\bibnamefont {Behnia}},\ }\bibfield  {title}
	{\bibinfo {title} {Phonon thermal hall effect in strontium titanate},\ }\href
	{https://doi.org/10.1103/PhysRevLett.124.105901} {\bibfield  {journal}
		{\bibinfo  {journal} {Phys. Rev. Lett.}\ }\textbf {\bibinfo {volume} {124}},\
		\bibinfo {pages} {105901} (\bibinfo {year} {2020})}\BibitemShut {NoStop}%
	\bibitem [{\citenamefont {Grissonnanche}\ \emph {et~al.}(2019)\citenamefont
		{Grissonnanche}, \citenamefont {Legros}, \citenamefont {Badoux},
		\citenamefont {Lefran{\c{c}}ois}, \citenamefont {Zatko}, \citenamefont
		{Lizaire}, \citenamefont {Lalibert{\'e}}, \citenamefont {Gourgout},
		\citenamefont {Zhou}, \citenamefont {Pyon}, \citenamefont {Takayama},
		\citenamefont {Takagi}, \citenamefont {Ono}, \citenamefont {Doiron-Leyraud},\
		and\ \citenamefont {Taillefer}}]{Grissonnanche2019Cuprate}%
	\BibitemOpen
	\bibfield  {author} {\bibinfo {author} {\bibfnamefont {G.}~\bibnamefont
			{Grissonnanche}}, \bibinfo {author} {\bibfnamefont {A.}~\bibnamefont
			{Legros}}, \bibinfo {author} {\bibfnamefont {S.}~\bibnamefont {Badoux}},
		\bibinfo {author} {\bibfnamefont {E.}~\bibnamefont {Lefran{\c{c}}ois}},
		\bibinfo {author} {\bibfnamefont {V.}~\bibnamefont {Zatko}}, \bibinfo
		{author} {\bibfnamefont {M.}~\bibnamefont {Lizaire}}, \bibinfo {author}
		{\bibfnamefont {F.}~\bibnamefont {Lalibert{\'e}}}, \bibinfo {author}
		{\bibfnamefont {A.}~\bibnamefont {Gourgout}}, \bibinfo {author}
		{\bibfnamefont {J.-S.}\ \bibnamefont {Zhou}}, \bibinfo {author}
		{\bibfnamefont {S.}~\bibnamefont {Pyon}}, \bibinfo {author} {\bibfnamefont
			{T.}~\bibnamefont {Takayama}}, \bibinfo {author} {\bibfnamefont
			{H.}~\bibnamefont {Takagi}}, \bibinfo {author} {\bibfnamefont
			{S.}~\bibnamefont {Ono}}, \bibinfo {author} {\bibfnamefont {N.}~\bibnamefont
			{Doiron-Leyraud}},\ and\ \bibinfo {author} {\bibfnamefont {L.}~\bibnamefont
			{Taillefer}},\ }\bibfield  {title} {\bibinfo {title} {{Giant thermal Hall
				conductivity in the pseudogap phase of cuprate superconductors}},\
	}\href@noop {} {\bibfield  {journal} {\bibinfo  {journal} {Nature}\ }\textbf
		{\bibinfo {volume} {571}},\ \bibinfo {pages} {376} (\bibinfo {year}
		{2019})}\BibitemShut {NoStop}%
	\bibitem [{\citenamefont {Ando}\ \emph {et~al.}(1995)\citenamefont {Ando},
		\citenamefont {Boebinger}, \citenamefont {Passner}, \citenamefont {Kimura},\
		and\ \citenamefont {Kishio}}]{Ando1995}%
	\BibitemOpen
	\bibfield  {author} {\bibinfo {author} {\bibfnamefont {Y.}~\bibnamefont
			{Ando}}, \bibinfo {author} {\bibfnamefont {G.~S.}\ \bibnamefont {Boebinger}},
		\bibinfo {author} {\bibfnamefont {A.}~\bibnamefont {Passner}}, \bibinfo
		{author} {\bibfnamefont {T.}~\bibnamefont {Kimura}},\ and\ \bibinfo {author}
		{\bibfnamefont {K.}~\bibnamefont {Kishio}},\ }\bibfield  {title} {\bibinfo
		{title} {{Logarithmic Divergence of both In-Plane and Out-of-Plane
				Normal-State Resistivities of Superconducting La$_{2-x}$Sr$_x$CuO$_4$ in the
				Zero-Temperature Limit}},\ }\href
	{https://doi.org/10.1103/PhysRevLett.75.4662} {\bibfield  {journal} {\bibinfo
			{journal} {Phys. Rev. Lett.}\ }\textbf {\bibinfo {volume} {75}},\ \bibinfo
		{pages} {4662} (\bibinfo {year} {1995})}\BibitemShut {NoStop}%
	\bibitem [{\citenamefont {Ito}\ \emph {et~al.}(2013)\citenamefont {Ito},
		\citenamefont {Ushiyama}, \citenamefont {Yanagisawa}, \citenamefont
		{Tomioka}, \citenamefont {Shindo},\ and\ \citenamefont
		{Yanase}}]{ITO2013264}%
	\BibitemOpen
	\bibfield  {author} {\bibinfo {author} {\bibfnamefont {T.}~\bibnamefont
			{Ito}}, \bibinfo {author} {\bibfnamefont {T.}~\bibnamefont {Ushiyama}},
		\bibinfo {author} {\bibfnamefont {Y.}~\bibnamefont {Yanagisawa}}, \bibinfo
		{author} {\bibfnamefont {Y.}~\bibnamefont {Tomioka}}, \bibinfo {author}
		{\bibfnamefont {I.}~\bibnamefont {Shindo}},\ and\ \bibinfo {author}
		{\bibfnamefont {A.}~\bibnamefont {Yanase}},\ }\bibfield  {title} {\bibinfo
		{title} {Laser-diode-heated floating zone (ldfz) method appropriate to
			crystal growth of incongruently melting materials},\ }\href
	{https://doi.org/https://doi.org/10.1016/j.jcrysgro.2012.10.059} {\bibfield
		{journal} {\bibinfo  {journal} {Journal of Crystal Growth}\ }\textbf
		{\bibinfo {volume} {363}},\ \bibinfo {pages} {264} (\bibinfo {year}
		{2013})}\BibitemShut {NoStop}%
	\bibitem [{\citenamefont {Viskadourakis}\ \emph {et~al.}(2015)\citenamefont
		{Viskadourakis}, \citenamefont {Sunku}, \citenamefont {Mukherjee},
		\citenamefont {Andersen}, \citenamefont {Ito}, \citenamefont {Sasagawa},\
		and\ \citenamefont {Panagopoulos}}]{Viskadourakis2015Cuprates}%
	\BibitemOpen
	\bibfield  {author} {\bibinfo {author} {\bibfnamefont {Z.}~\bibnamefont
			{Viskadourakis}}, \bibinfo {author} {\bibfnamefont {S.}~\bibnamefont
			{Sunku}}, \bibinfo {author} {\bibfnamefont {S.}~\bibnamefont {Mukherjee}},
		\bibinfo {author} {\bibfnamefont {B.}~\bibnamefont {Andersen}}, \bibinfo
		{author} {\bibfnamefont {T.}~\bibnamefont {Ito}}, \bibinfo {author}
		{\bibfnamefont {T.}~\bibnamefont {Sasagawa}},\ and\ \bibinfo {author}
		{\bibfnamefont {C.}~\bibnamefont {Panagopoulos}},\ }\bibfield  {title}
	{\bibinfo {title} {Ferroelectricity in underdoped la-based cuprates},\
	}\href@noop {} {\bibfield  {journal} {\bibinfo  {journal} {Sci. Rep.}\
		}\textbf {\bibinfo {volume} {5}},\ \bibinfo {pages} {15268} (\bibinfo {year}
		{2015})}\BibitemShut {NoStop}%
	\bibitem [{\citenamefont {Fujita}\ \emph {et~al.}(2011)\citenamefont {Fujita},
		\citenamefont {Hiraka}, \citenamefont {Matsuda}, \citenamefont {Matsuura},
		\citenamefont {M.~Tranquada}, \citenamefont {Wakimoto}, \citenamefont {Xu},\
		and\ \citenamefont {Yamada}}]{Fujita2011Cuprates}%
	\BibitemOpen
	\bibfield  {author} {\bibinfo {author} {\bibfnamefont {M.}~\bibnamefont
			{Fujita}}, \bibinfo {author} {\bibfnamefont {H.}~\bibnamefont {Hiraka}},
		\bibinfo {author} {\bibfnamefont {M.}~\bibnamefont {Matsuda}}, \bibinfo
		{author} {\bibfnamefont {M.}~\bibnamefont {Matsuura}}, \bibinfo {author}
		{\bibfnamefont {J.}~\bibnamefont {M.~Tranquada}}, \bibinfo {author}
		{\bibfnamefont {S.}~\bibnamefont {Wakimoto}}, \bibinfo {author}
		{\bibfnamefont {G.}~\bibnamefont {Xu}},\ and\ \bibinfo {author}
		{\bibfnamefont {K.}~\bibnamefont {Yamada}},\ }\bibfield  {title} {\bibinfo
		{title} {{Progress in neutron scattering studies of spin excitations in
				high-T$_c$ cuprates}},\ }\href@noop {} {\bibfield  {journal} {\bibinfo
			{journal} {Phys. Soc. Jpn.}\ }\textbf {\bibinfo {volume} {81}},\ \bibinfo
		{pages} {011007} (\bibinfo {year} {2011})}\BibitemShut {NoStop}%
\end{thebibliography}

\onecolumngrid
\newpage
\setcounter{section}{0}
\setcounter{figure}{0}
\renewcommand{\figurename}{{ FIG.}}
\renewcommand{\tablename}{{TABLE}}
\renewcommand{\theequation}{S\arabic{equation}}
\renewcommand{\thefigure}{S\arabic{figure}}
\renewcommand{\thetable}{S\arabic{table}}
\renewcommand{\bibnumfmt}[1]{[S#1]}
\renewcommand{\citenumfont}[1]{S#1}

\begin{center}
	\textbf{Supplemental Material for ``Thermal Hall Resistivity as a Unifying Description of \\ Phonon Thermal Hall Effect in Various Insulators''}\\
	\fontsize{9}{12}\selectfont
	\vspace{2em}
	Dirui Wu,$^{1}$ Ying Kit Tsui,$^{1}$ Xavier Loh,$^{1}$ Minseong Lee,$^{2}$ Eun Sang Choi,$^{3}$ Takao Sasagawa,$^{4}$ Toshimitsu Ito,$^{5}$ \\ Xiao Renshaw Wang,$^{1,6}$ Pinaki Sengupta,$^{1}$ Xinyang Zhang,$^{1,*}$ and Christos Panagopoulos$^{1,\dagger}$\\
	\vspace{1em}
	$^1${\it Division of Physics and Applied Physics, School of Physical and Mathematical Sciences, \\Nanyang Technological University, Singapore 637371, Singapore}\\
	$^2${\it National High Magnetic Field Laboratory, Los Alamos National Laboratory, Los Alamos, New Mexico 87545, USA}\\
	$^3${\it National High Magnetic Field Laboratory, Tallahassee, Florida 32310, USA}\\
	$^4${\it Materials and Structures Laboratory, Institute of Science Tokyo, Yokohama, Kanagawa 226-8501, Japan}\\
	$^5${\it National Institute of Advanced Industrial Science and Technology, Tsukuba, Ibaraki 305-8565, Japan}\\
	$^6${\it School of Electrical and Electronic Engineering, \\ Nanyang Technological University, Singapore 639798, Singapore}\\
\end{center}

\vspace{1.5cm}

\textbf{\MakeUppercase{S1. Thermal Hall methodology for LSNO-1/3}}
\vspace{0.5cm}

\textbf{\MakeUppercase{S2. Electrical and Magnetic Characterization for LSNO-1/3}}
\vspace{0.5cm}

\textbf{\MakeUppercase{S3. Calculation of $\bm{\kappa_{xy}(H)}$ using in-situ calibrated $\bm{\kappa_{xx}(H)}$}}
\vspace{0.5cm}

\textbf{\MakeUppercase{S4. Power-law divergence of $\bm{w_{xx}(T)}$ and power-law of $\bm{\kappa_{xy}(T)}$}}
\vspace{0.5cm}

\textbf{\MakeUppercase{S5. Diverging temperature dependence of thermal Hall coefficient}}
\vspace{0.5cm}

\textbf{\MakeUppercase{S6. $\bm{T}$-linear thermal Hall angle in STO, S\lowercase{i}, and G\lowercase{e}}}
\vspace{0.5cm}

\textbf{\MakeUppercase{S7. Logarithmic divergence of $\bm{w_{xy}(T)}$ in LSCO-0.06}}
\vspace{0.5cm}

\textbf{\MakeUppercase{S8. Supporting data:\newline
	\hspace*{0.945cm} Thermal Hall methodology and characterization for LSCO-0.001}}
\vspace{0.5cm}

\textbf{\MakeUppercase{S9. Supporting data:\newline \hspace*{0.945cm} Thermal Hall effect in LSCO-0.001}}
\vspace{0.5cm}

\textbf{\MakeUppercase{S10. Extended thermal Hall measurement of LSNO-1/3 at higher \newline \hspace*{1.15cm} temperatures}}
\newpage

\section{S1. Thermal Hall methodology for LSNO-1/3}
\begin{figure}[h]\centering
\resizebox{18cm}{!}{
	\includegraphics{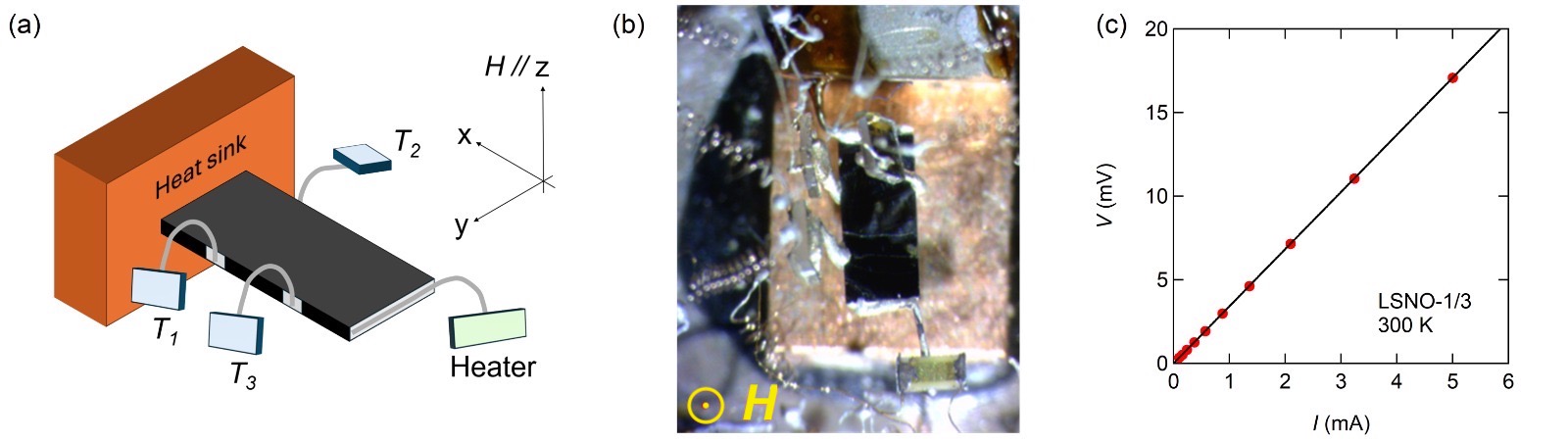}
}       
\caption{(a)~Schematic for measuring the thermal Hall effect in LSNO-1/3. The heater generates a heat current in the x-direction, which aligned with the $a$- or $b$-axis of the sample. Magnetic field is applied perpendicular to the xy-plane. The three bare-chip Cernox thermometers measure temperatures $T_1$, $T_2$, and $T_3$. (b)~Photograph of the measurement setup for LSNO-1/3. (c)~The current-voltage ($I$-$V$) characteristic of two contact pads on LSNO-1/3 at 300 K.
	\label{fig:S1}
}
\end{figure}

Single-crystal \LSNOthird\ (LSNO-1/3) was grown using the floating zone method. Figure~\ref{fig:S1}(a) and (b) show the schematic and the photograph, respectively, of the measurement setup for LSNO-1/3. For thermal transport measurement, the sample was first shaped into a 2.6 mm $\times$ 1.1 mm $\times$ 0.2 mm slab, using a wire saw and a crystal polisher with 9 $\mu m$ diamond lapping film. Then, to establish ohmic contacts with the quasi-2d Ni-O layers, contact pads were made on the sample's side surfaces using DuPont 6838 silver paste followed by annealing under oxygen flow of 100cc/min in a tube furnace at temperatures presented in Table~\ref{tab:AnnealingSeq}. The oxygen flow minimized the changing of the oxygen content in the sample at high temperature. Figure~\ref{fig:S1}(c) presents the contacts' current-voltage ($I$-$V$) characteristics measured across two contact pads on LSNO-1/3 using pseudo-four-point method. The voltage increases linearly with the current, supporting that the contacts are ohmic. The two contact pads along the width of the sample are for the heater and the heat sink. The three contact pads along the length of the sample are for the Cernox thermometers, with two aligned on the opposite sides for the thermal Hall measurement. All the components were attached to the corresponding contact pads using silver epoxy and 100-$\mu$m silver wires.

\begin{table}[htbp]
\centering
\caption{Annealing sequence for LSNO-1/3}
\label{tab:AnnealingSeq}
\vspace{0.2cm}
\begin{tabular*}{0.8\linewidth}{@{\extracolsep{\fill}} ccc @{}}
	\hline\hline\noalign{\smallskip}
	Temperature Ramp Rate ($^\circ$C/hr) & Temperature Setpoint ($^\circ$C) & Dwell time (min) \\
	\noalign{\smallskip}\hline\noalign{\smallskip}
	150 & 400 & 0  \\
	50  & 450 & 10 \\
	30  & 20  & 0  \\
	\noalign{\smallskip}\hline\hline
\end{tabular*}
\end{table}

The heater and each Cernox thermometer were measured using thin manganin wires in pseudo-four-point configuration. The thin wires establish electrical connections while minimizing heat leakage. The electrical resistances of the Cernox thermometers were measured using a resistance bridge (Lakeshore 372 and 3726 scanner). For calibration, the electrical resistance $R(T,H)$ of each Cernox thermometer were measured with heater off at thermal equilibrium at different temperatures $T$ at different magnetic fields $H$ against pre-calibrated thermometers. The field correction of the pre-calibrated thermometer was done by $^3$He vapor pressure thermometry and capacitance temperature sensor below and above 1.7 K, respectively. Thus, the temperatures $T_1$, $T_2$, and $T_3$ could be derived from the corresponding measured resistances.

With heater current along the x-direction, the longitudinal temperature difference ($\Delta T_x = T_3 - T_1$) and transverse temperature difference ($\Delta T_y = T_2 - T_1$) across the sample were measured by sweeping temperature at fixed magnetic field perpendicular to the xy-plane. Here, the sample was positioned so that the $a$- or $b$-axis aligns with the $x$-direction and the $c$-axis is perpendicular to the xy-plane. While sweeping temperature, the heater power decreased linearly with decreasing temperature. Due to small unavoidable misalignments of the contacts, $\Delta T_x$ and $\Delta T_y$ were measured under both positive and negative magnetic field, so that they could be symmetrized and anti-symmetrized respectively. 

From Fourier's law, we have

\begin{equation}
\begin{bmatrix}
	\Delta T_{x} / l_x \\ \Delta T_{y} / l_y 
\end{bmatrix}
=-
\begin{bmatrix}
	w_{xx} & w_{xy} \\ w_{yx} & w_{yy}
\end{bmatrix}
\begin{bmatrix}
	P/(tl_y) \\ 0
\end{bmatrix},
\label{eq:S1}
\end{equation}
\newline
where $l_x$ is the separation between the $T_2$ and $T_3$ contact pads, $l_y$ is the separation between the $T_1$ and $T_2$ contact pads, $t$ is the thickness of the sample, $P$ is the heater power, $w_{xx}$ is the longitudinal thermal resistivity, and $w_{xy}$ is the thermal Hall resistivity. Since $w_{xx} = w_{yy}$ for in-plane isotropy in LSNO-1/3~\cite{Klingeler2005} and $w_{xy} = - w_{yx}$ for Onsager relations, Eq.~\ref{eq:S1} simplifies to

\begin{align}
w_{xx} = -\frac{\Delta T_{x} tl_y}{l_xP}, \quad \quad \quad \quad
w_{xy} = \frac{\Delta T_{y}t}{P}.  
\label{eq:S2}
\end{align}
\newline
Therefore, the thermal resistivity tensor is a directly measured quantity. One the other hand, as the thermal conductivity tensor is the inverse of the thermal resistivity tensor, \ie $\bm{\kappa}=\mathbf{w}^{-1}$, the longitudinal thermal conductivity $\kappa_{xx}$ and thermal Hall conductivity $\kappa_{xy}$ are calculated by

\begin{equation}
\kappa_{xx} = \frac{w_{xx}}{w_{xx}^2 + w_{xy}^2} \approx \frac{1}{w_{xx}}, \quad \quad \quad \quad \kappa_{xy} = \frac{-w_{xy}}{w_{xx}^2 + w_{xy}^2} \approx -\frac{w_{xy}}{w_{xx}^2},
\label{eq:S3}
\end{equation}
where the simplifications are justified by $w_{xy}/w_{xx} \sim 0.001$. The inclusion of both $w_{xx}$ and $w_{xy}$ to calculate $\kappa_{xy}$ implies that any analysis of $\kappa_{xy}$ incorporates the longitudinal and transverse temperature differences. 
\newpage

\section{S2. Electrical and Magnetic Characterization for LSNO-1/3}
\begin{figure}[h]\centering
\resizebox{12cm}{!}{
	\includegraphics{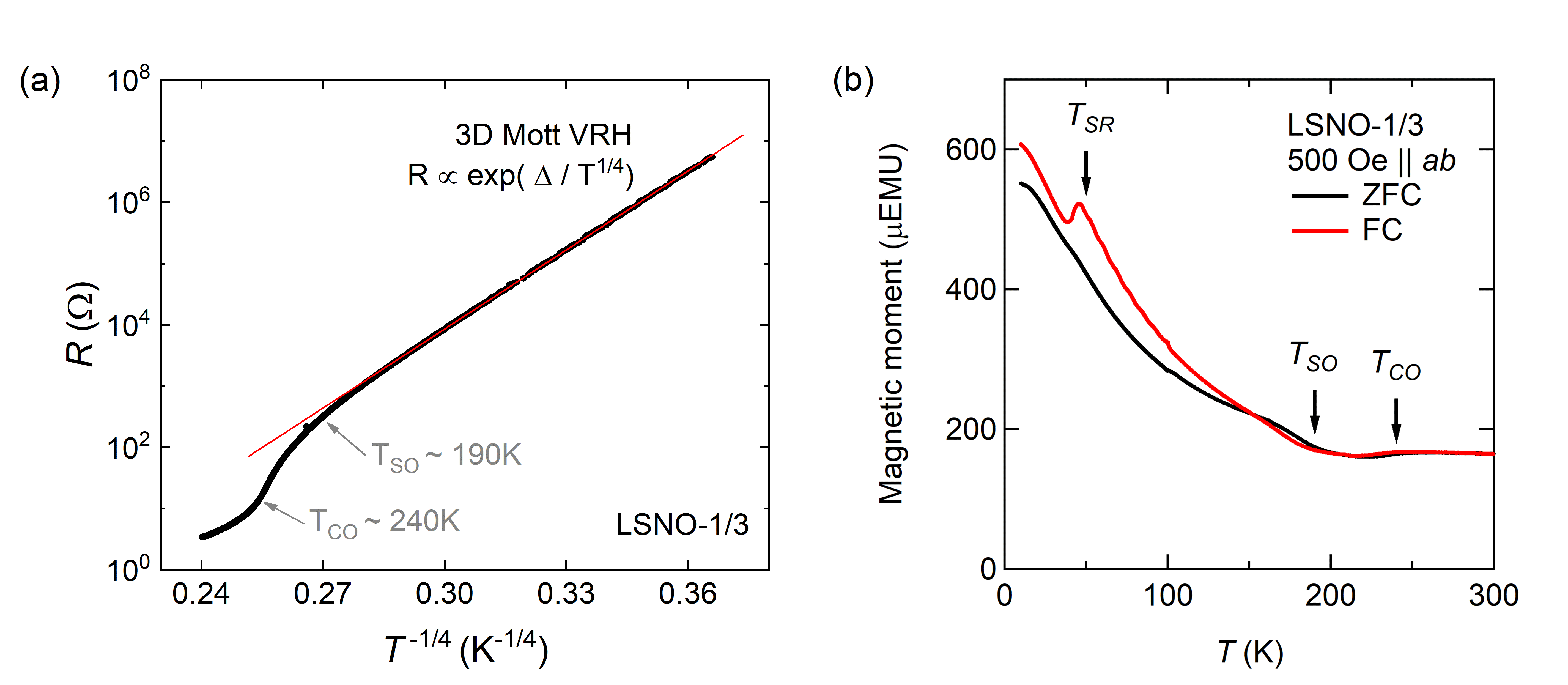}
}       
\caption{(a)~Temperature dependence of resistance of LSNO-1/3. (b)~Zero-field-cooled (ZFC) and field-cooled (FC) magnetic moments of LSNO-1/3 measured with magnetic field of 500 Oe along the $ab$-plane. 
	\label{fig:S2}
}
\end{figure}

\LSNOthird\ (LSNO-1/3) is a Ruddlesden-Popper layered perovskite~\cite{55_Freeman2006} and remains in the tetragonal structure at all measured temperatures~\cite{Hucker2004}. Substitution of Sr$^{2+}$ for La$^{3+}$ in LSNO-1/3 introduces holes to the NiO layers, resulting in a mixed valence state of Ni$^{3+}$ ([Ar]$3d^7$) and Ni$^{2+}$ ([Ar]$3d^8$). Within the temperature range of our study $T \leq 28 K$, glassy dynamics dominate the stripe orders~\cite{53_Park2005,54_Spencer2005,55_Freeman2006,56_Filippi2009} along with strong electron-phonon coupling~\cite{47_CHChen1993,57_McQueeney1999}, quasi-1D antiferromagnetic (AFM) magnons~\cite{Merritt2019, 58_Freeman2011}, and local octahedral distortion~\cite{50_Zachar1998}. Figure~\ref{fig:S2}(a) shows the temperature dependence of the electrical resistance of LSNO-1/3. The low-temperature behavior exhibits Mott variable-range hopping, suggesting carrier localization in this doped insulator~\cite{MATSUSHITA1990}. The two inflexion points correspond to charge and spin stripe ordering temperatures, $T_\text{CO}$ and $T_\text{SO}$ as indicated by the arrows~\cite{Hess1999}.

Figure~\ref{fig:S2}(b) shows the temperature dependence of the magnetic moment of LSNO-1/3 measured using the vibrating sample magnetometer (VSM) option in a Physical Property Measurement System (PPMS). Magnetic field of 500 Oe was applied along the $ab$-plane. The clear splitting zero-field-cooled~(ZFC) and field-cooled~(FC) magnetic moment is consistent with the previous report~\cite{55_Freeman2006}. The characteristic temperatures $T_\text{CO} \sim 240~\text{K}$ and $T_\text{SO} \sim 190~\text{K}$, established from neutron scattering and diffraction studies~\cite{47_CHChen1993, 51_Yoshizawa2000, Anissimova2014LSNO, Lee2001LSNO}, correspond to the onset of charge-stripe and spin-stripe ordering respectively. The feature at $T_\text{SR}\sim50~\text{K}$ is indicative of a spin reorientation transition.
\newpage

\section{S3. Calculation of $\boldsymbol{\kappa_{xy}(H)}$ using in-situ calibrated $\boldsymbol{\kappa_{xx}(H)}$}
\begin{figure}[h]\centering
\resizebox{12cm}{!}{
	\includegraphics{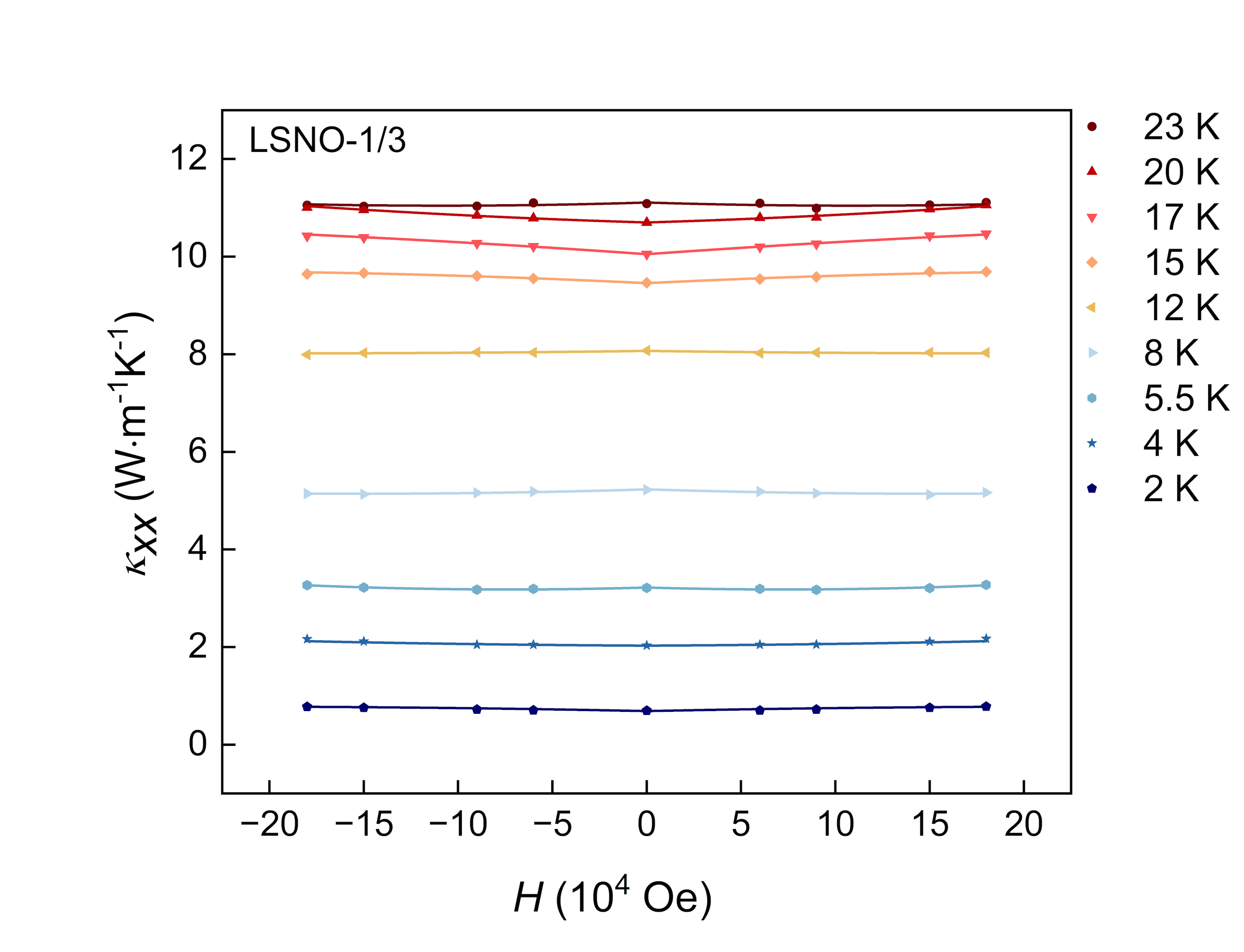}
}       
\caption{Field dependence of longitudinal thermal conductivity $\kappa_{xx}$ of LSNO-1/3 at various temperatures from 2 K to 23 K, as extracted from fixed-field temperature-sweeps.
	\label{fig:S3}
}
\end{figure}

To obtain the thermal Hall conductivity $\kappa_{xy}$, we need both the longitudinal temperature difference $\Delta T_x$ and the transverse temperature difference $\Delta T_y$. To calculate the temperature dependence of $\kappa_{xy}$ of \LSNOthird\ (LSNO-1/3), we used the $\Delta T_y$ and $\Delta T_x$ measured by sweeping temperature at fixed magnetic fields. On the other hand, to calculate the field dependence of $\kappa_{xy}$ of LSNO-1/3, we used the $\Delta T_y$ measured by sweeping field at fix temperatures but the $\Delta T_x$ measured by sweeping temperature at fixed magnetic fields, because the \textit{in situ} calibration is more accurate for temperature sweeps.

Here, we explain how the $\Delta T_x(H)$ was extracted from the temperature sweeps at fixed fields. Figure~\ref{fig:S3} shows the field dependence of the longitudinal thermal conductivity  $\kappa_{xx}$ of LSNO-1/3 up to our highest field of $18\times10^4$~Oe, at different temperatures from 2 K to 23 K. These datasets were extracted from the data presented in Fig.~1(c) in the main text. For all temperatures, the $\kappa_{xx}$ at different fields can be well fitted by a 2nd-order even polynomial function $f(x)=c_0+c_1|x|+c_2x^2$. Since $\kappa_{xx}\approx 1/w_{xx}$ and $w_{xx}$ is related to $\Delta T_x$ by Eq.~\ref{eq:S2}, from the fittings, we have effectively obtained the $\Delta T_x(H)$ required. The $\kappa_{xy}(H)$ was obtained by

\begin{equation}
\kappa_{xy}=\frac{\Delta T_y \cdot t \cdot \kappa_{xx}^2}{P},
\end{equation}
\newline
where $t$ is the sample thickness and $P$ is the heater power.
\newpage

\section{S4. Power-law divergence of $\boldsymbol{w_{xx}(T)}$ and power-law of $\boldsymbol{\kappa_{xy}(T)}$}
\begin{figure}[h]\centering
\resizebox{18cm}{!}{
	\includegraphics{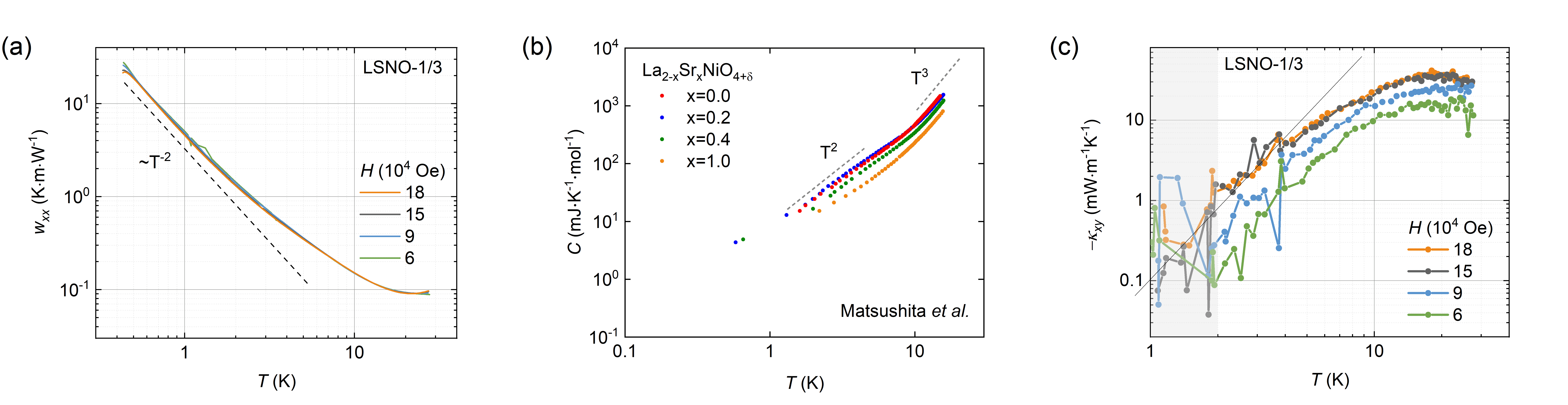}
}       
\caption{(a)~Temperature dependence of the thermal resistivity $w_{xx}$ of LSNO-1/3 in log-log scale. (b)~Temperature dependence of the specific heat capacity of LSNO in log-log scale extracted from Ref.~\cite{MATSUSHITA1990}. (c)~Temperature dependence of the thermal Hall conductivity $\kappa_{xy}$ of LSNO-1/3 in log-log scale.
	\label{fig:S4}
}
\end{figure}

The temperature dependence of the longitudinal thermal resistivity $w_{xx}$ of LSNO-1/3 at various magnetic fields is shown in log-log scale in Fig.~\ref{fig:S4}(a). At low temperatures below $\sim$2~K, $w_{xx}(T)$ follows a $T^{-2}$ power-law divergence. The temperature dependence of the specific heat of LSNO from Ref.~\cite{MATSUSHITA1990} is plotted in log-log scale in Fig.~\ref{fig:S4}(b), which highlights the power-law scaling exponents. As the temperature decreases, the specific heat capacity initially follows a $T^3$ dependence and then a $T^2$ dependence at lower temperature. From the kinetic theory of phonon transport, $\kappa \sim Cvl/3$, where $\kappa$ is the thermal conductivity, $C$ is the phonon specific heat, $v$ the sound velocity, and $l$ the phonon mean free path. The variation in $v$ of LSNO-1/3 is small across the temperature range of 150 to 300~K~\cite{Ramirez1996}. Therefore, assuming $v$ does not vary significantly at low temperature, the phonon mean free path $l$ remains approximately weakly temperature dependent below $\sim$2 K. The $T^2$ power law in $C$ is in contrast with the conventional three-dimensional Debye phonon behavior ($\propto T^3$), and indicates a reduction of the heat-carrying phonon excitations. Above $\sim$2~K, $w_{xx}$ deviates from the $T^{-2}$ power-law and develops a minimum around $\sim$20~K, coinciding with the intermediate-temperature peak in $\kappa_{xx}$ as discussed in the main text.

Figure~1(a) of the main text shows the temperature dependence of the thermal Hall conductivity $\kappa_{xy}$ at various fields. Figure~\ref{fig:S4}(c) shows the same plot but in log-log scale. For all measured magnetic fields, as the temperature increases from the base temperature, the magnitude of $\kappa_{xy}$ follows an approximate $T^3$ power-law (solid line) before reaching a peak around $\sim$20~K. Combining the approximate $\kappa_{xy} \propto T^3$ behavior with the $\kappa_{xx} \propto T^2$ power-law established in Fig.~1(c) in the main text, the thermal Hall angle $\kappa_{xy}/\kappa_{xx}$ follows a $T$-linear behavior at low temperatures, as demonstrated in Fig.~2(c) of the main text. 
\newpage

\section{S5. Diverging temperature dependence of thermal Hall coefficient}
\begin{figure}[h]\centering
\resizebox{12cm}{!}{
	\includegraphics{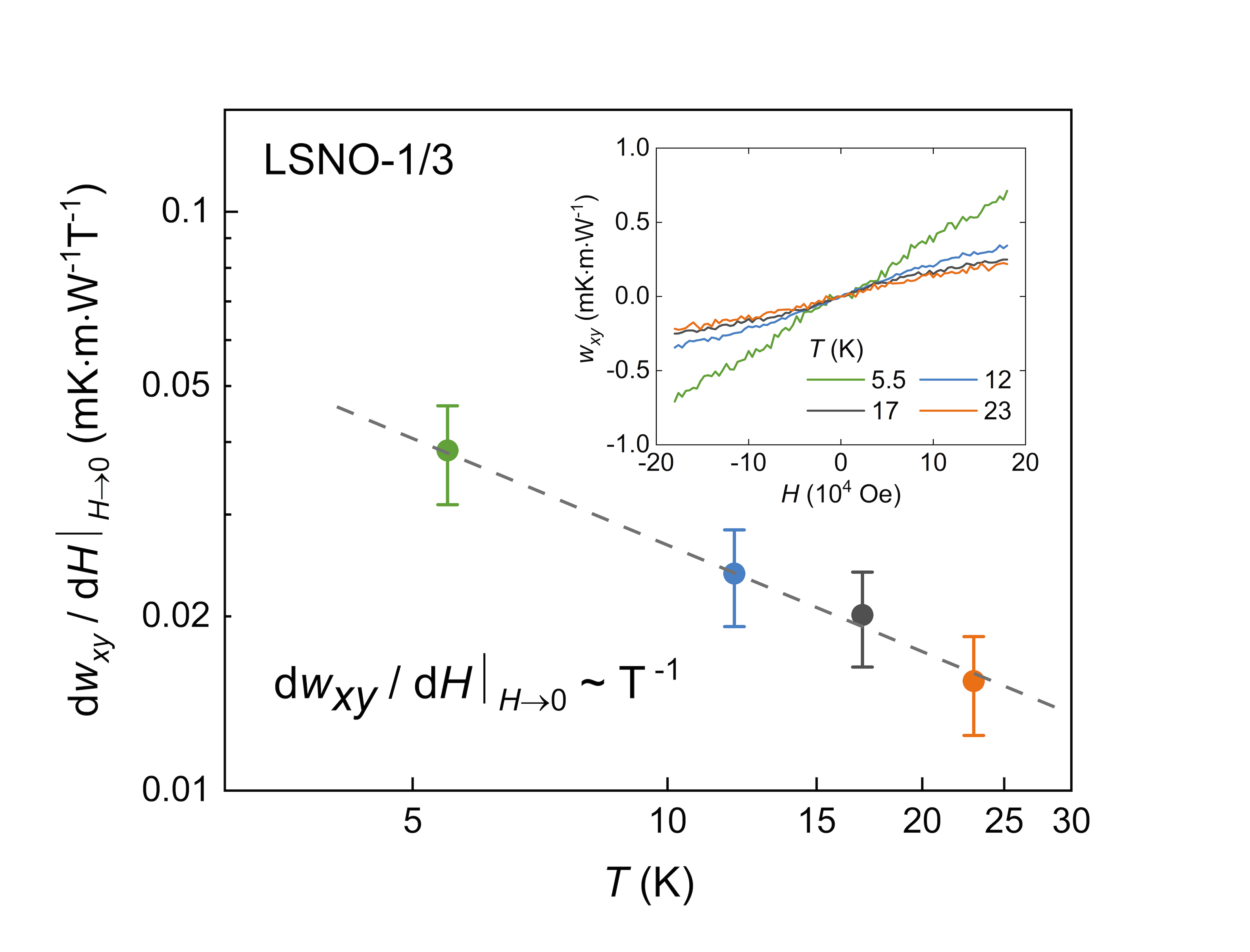}
}       
\caption{The thermal Hall coefficient of LSNO-1/3 at the zero-field limit at various temperatures in log-log scale. The inset shows the field dependence of the thermal Hall resistivity $w_{xy}$ from which the thermal Hall coefficients were extracted.
	\label{fig:S5}
}
\end{figure}

To better indicate the diverging temperature dependence of the thermal Hall resistivity of \LSNOthird\ (LSNO-1/3) at low fields, we present the thermal Hall coefficient at various temperatures in Fig.~\ref{fig:S5}. Here, thermal Hall coefficient is given by
\begin{equation}
R_{TH} = \frac{dw_{xy}}{dH}\bigg|_{H=0} = -\frac{d(\kappa_{xy}/\kappa_{xx}^2)}{dH}\bigg|_{H=0},
\end{equation}
where $w_{xy}$ is the thermal Hall resistivity, $\kappa_{xy}$ is the thermal Hall conductivity, and $\kappa_{xx}$ is the longitudinal thermal conductivity. Since $w_{xy}$ is directly proportional to the transverse temperature difference, $R_{TH}$ quantifies how strongly the heat flow is deflected into the transverse direction per unit applied magnetic field in the zero-field limit. In Fig.~\ref{fig:S5}, the $R_{TH}$ is extracted from the low-field linear region of the $w_{xy}$ field dependence shown in the inset. This inset is showing the same datasets as in Fig. 2(b) in the main text. The extracted $R_{TH}(T)$ follows a $T^{-1}$ power-law dependence (dashed line), demonstrating that the insulating-like divergence of $w_{xy}$ persists in the zero-field limit, consistent with the $T^{-1}$ divergence of $w_{xy}(T)$ at higher fields up to 18~T. 
\newpage

\section{S6. $\boldsymbol{T}$-linear thermal Hall angle in STO, S\lowercase{i}, and G\lowercase{e}}
\begin{figure}[h]\centering
\resizebox{12cm}{!}{
	\includegraphics{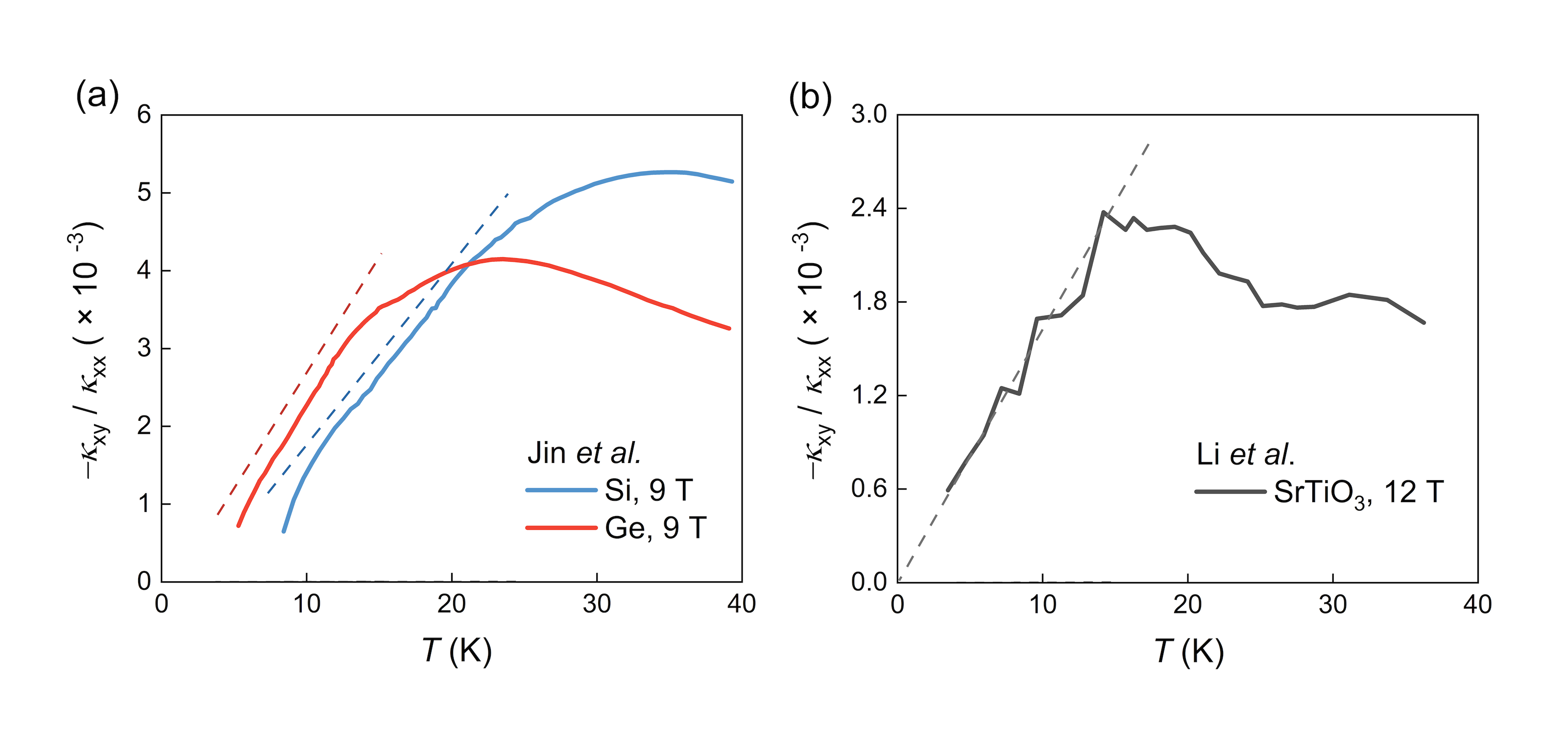}
}       
\caption{(a)~The temperature dependence of thermal Hall angles $-\kappa_{xy}/\kappa_{xx}$ of Si and Ge from Ref.~\cite{Jin2025Silicon}. (b)~The temperature dependence of thermal Hall angle $-\kappa_{xy}/\kappa_{xx}$ of SrTiO$_3$ (STO) from Ref.~\cite{XiaoKang2020}.
	\label{fig:S6}
}
\end{figure}

To demonstrate the $T$-linear behavior of the thermal Hall angle as a unifying phenomenon in various insulators, we have plotted Fig.~\ref{fig:S6}(a) and (b). Figure~\ref{fig:S6}(a) shows the thermal Hall angle as a function of temperature for semiconductors Si and Ge extracted from {\cite{Jin2025Silicon}, and Fig.~\ref{fig:S6}(b) shows the thermal Hall angle as a function of temperature for SrTiO$_3$~(STO) extracted from {\cite{XiaoKang2020}. The thermal Hall angle can be obtained by dividing the negative of thermal Hall conductivity $\kappa_{xy}$ by longitudinal thermal conductivity $\kappa_{xx}$. For Si and Ge, their thermal Hall angles were calculated using $\kappa_{xx}(0~\text{T})$ due to data availability. For all three materials, the thermal Hall angle exhibits $T$-linear dependence at low temperatures, as indicated by the dashed lines, before forming a peak at higher temperatures.
	
The $T$-linear behavior of the thermal Hall angle at low temperature arises from the ratio of $\kappa_{xy}(T)$ to $\kappa_{xx}(T)$, where $\kappa_{xy}(T)$ and $\kappa_{xx}(T)$ can each have its own power-law exponent. For the $\kappa_{xy}(T)$ and $\kappa_{xx}(T)$ of Si and Ge~\cite{Jin2025Silicon}, their corresponding functional forms yield the $T$-linear behavior in the thermal Hall angle. The same applies to STO and \LSNOthird\ (LSNO-1/3), where $\kappa_{xx}(T)\propto T^3$ and $\kappa_{xy}(T)\propto T^4$ for STO~\cite{XiaoKang2020}, and $\kappa_{xx}(T)\propto T^2$ and $\kappa_{xy}(T)\propto T^3$ for LSNO-1/3. In all cases, the $T$-linear thermal Hall angle emerges as a result of $\kappa_{xy}$ carrying one additional power of $T$ compared to $\kappa_{xx}$, despite the different forms in $\kappa_{xy}(T)$ and $\kappa_{xx}(T)$ across different systems, supporting the observation of $T$-linear thermal Hall angle as discussed in the main text.
\newpage

\section{S7. Logarithmic divergence of $\boldsymbol{w_{xy}(T)}$ in LSCO-0.06}
\begin{figure}[h]\centering
	\resizebox{8cm}{!}{
		\includegraphics{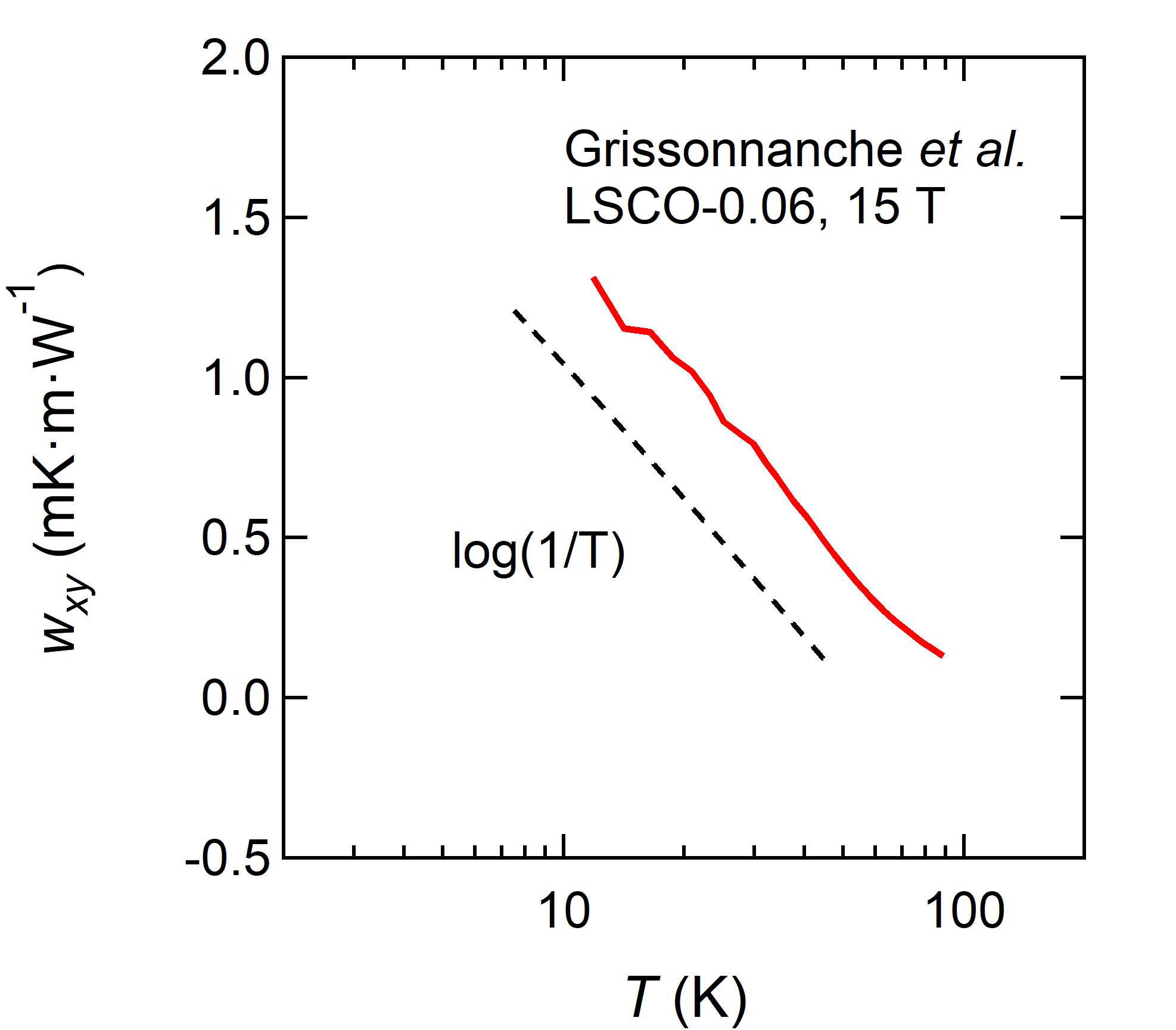}
	}       
	\caption{Temperature dependence of the thermal Hall resistivity $w_{xy}=-\kappa_{xy}/\kappa_{xx}^2$ of LSCO-0.06 at 15~T from Ref.~\cite{Grissonnanche2019Cuprate} plotted in linear-log scale.
		\label{fig:S7}
	}
\end{figure}
In Fig.~3(c) of the main text, the thermal Hall resistivity $w_{xy}$ of La$_{1.94}$Sr$_{0.06}$CuO$_4$ (LSCO-0.06) at 15 T extracted from Ref.~\cite{Grissonnanche2019Cuprate} is plotted as a function of temperature in log-log scale. To emphasize the logarithmic divergence behavior of $w_{xy}(T)$ in LSCO-0.06, we plot the same data set in linear-log scale in Fig.~\ref{fig:S7}. The $w_{xy}(T)$ is linear as expected, and can be represented by

\begin{equation}
	w_{xy} = a + b\log\frac{1\,\text{K}}{T} ,
\end{equation}
\newline
where $a = 2.85$~mK$\cdot$m$\cdot$W$^{-1}$ and 
$b = 1.42$~mK$\cdot$m$\cdot$W$^{-1}$.

The logarithmic divergence of $w_{xy}(T)$ in LSCO-0.06 is different from the $T^{-1}$ power-law divergence identified in \LSNOthird\ (LSNO-1/3) and La$_2$CuO$_4$ (LCO), and represents a weaker divergence. As discussed in the main text, the electrical resistivity of LSCO-0.06 also exhibits a comparable logarithmic divergence in the zero-temperature limit~\cite{Ando1995}, suggesting that the logarithmic divergence of $w_{xy}(T)$ in LSCO-0.06 may be connected to its localized electronic state. 
\newpage

\section{S8. Thermal Hall methodology and characterization for LSCO-0.001}
\begin{figure}[h]\centering
	\resizebox{18cm}{!}{
		\includegraphics{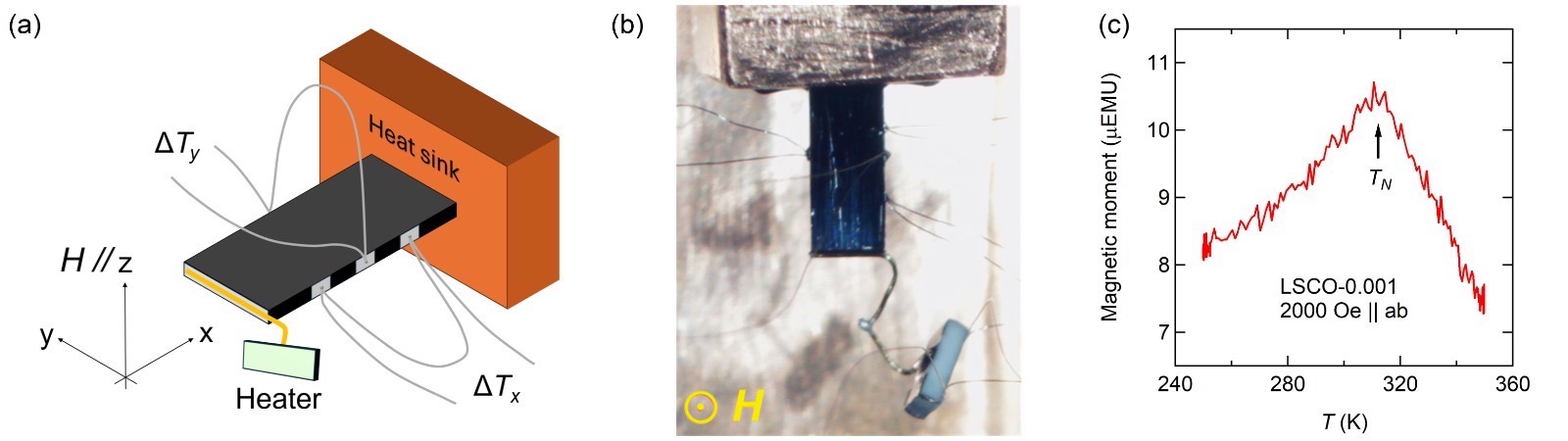}
	}       
	\caption{ (a)~Schematic for measuring the thermal Hall effect in LSCO-0.001. The heater generates a heat current in the x direction. The two thermocouples measure the longitudinal temperature difference $\Delta T_x$ and transverse temperature difference $\Delta T_y$. (b)~Photograph of the measurement setup for LSCO-0.001. (c)~The magnetic moment characteristic of LSCO-0.001.
		\label{fig:S8}
	}
\end{figure}

High-quality \LSCOthousandth\ (LSCO-0.001) single crystal was grown by the laser-diode-heated floating zone (LDFZ) method \cite{ITO2013264}. The as-grown sample with a N\'{e}el temperature ($T_N$) of 230 K was annealed in 0.1-Pa O$_2$ at 1000 $^\circ$C for 20 hours to remove the excess oxygen. The annealed sample showed a $T_N\approx 314$ K. Figure~\ref{fig:S8}(a) and (b) show the schematic and the photograph, respectively, of the thermocouple measurement setup for LSCO-0.001. The sample was shaped into a 0.27-mm-thick slab using a wire saw and a homemade crystal polisher with 9 $\mu m$ diamond lapping film. Annealed silver paste contact pads were made on the sample sides using the same procedures as that for the \LSNOthird\ (LSNO-1/3). The two contact pads along the width of the sample were for the heater and the heat sink. The four contact pads along the length of the sample were for the type-E differential thermocouples, with two pads aligned on the opposite sides for the thermal Hall measurement. The ratio of the width of the sample to the separation of the contact pads for $\Delta T_x$ is 1.1.

Using Stycast 2850 FT, we attached the heat sink, the junctions of the two type-E differential thermocouples, and the heater (through 100~$\mu$m-diameter gold wires connected by silver epoxy) to the corresponding contact pads on the sample. They were attached in such a way that the heat current flew in the x-direction, which aligns with the $a$-axis of LSCO-0.001, and the two type-E differential thermocouples measured the temperature differences $\Delta T_x$ and $\Delta T_y$. Here, a type-E differential thermocouple was made by spot welding the two ends of a 25~$\mu$m-diameter chromel wire to two 25~$\mu$m-diameter constantan wires to create two junctions. The two leads of the differential thermocouple were thermally anchored to a sapphire plate on the copper heat sink. Thus, the temperature difference across the two junctions are directly proportional to the voltage across the differential thermocouple. In contrast, the Cernox thermometers that we used for measuring the LSNO-1/3 measure the actual temperatures on the sample. To send electrical current to the heater and measure the voltage across, four 25~$\mu$m-diameter manganin wires were attached to the heater in pseudo-four-point configuration.

The high magnetic field up to 12~T and the low temperature down to 15~K for the thermal Hall measurements were achieved using a Physical Property Measurement System (PPMS). The sample was aligned such that the magnetic field is perpendicular to the xy-plane (see Fig.~\ref{fig:S8}) along the $c$-axis of LSCO-0.001. After the magnetic field and temperature reach the set points, we measured the voltage across each differential thermocouple while the heater was repeatedly turned ON for 80~s and then OFF for 80~s. This allowed us to suppress the voltage drift in the analysis. We calculate the voltage difference between ON and OFF for each repetition and calculate the average. The temperature differences $\Delta T_x$ and $\Delta T_y$ are obtained by dividing the corresponding average voltage differences by the Seebeck coefficient. Owing to small unavoidable misalignments of the contacts, $\Delta T_x$ and $\Delta T_y$ are symmetrized and anti-symmetrized respectively with respect to field.

Figure~\ref{fig:S8}(c) shows the temperature dependence of the magnetic moment of LSCO-0.001 under 2000~Oe along the $ab$-plane using the vibrating sample magnetometer (VSM) option in a PPMS. The magnetic moment shows a peak near 310~K. This peak is attributed to the N\'eel transition temperature $T_N$ of LSCO-0.001~\cite{Viskadourakis2015Cuprates}. It confirms that our LSCO-0.001 sample exhibits long-range antiferromagnetism without the charge/spin-ordered stripes~\cite{Fujita2011Cuprates}.
\newpage

\section{S9. Thermal Hall effet in LSCO-0.001}
\begin{figure}[h]\centering
	\resizebox{15 cm}{!}{
		\includegraphics{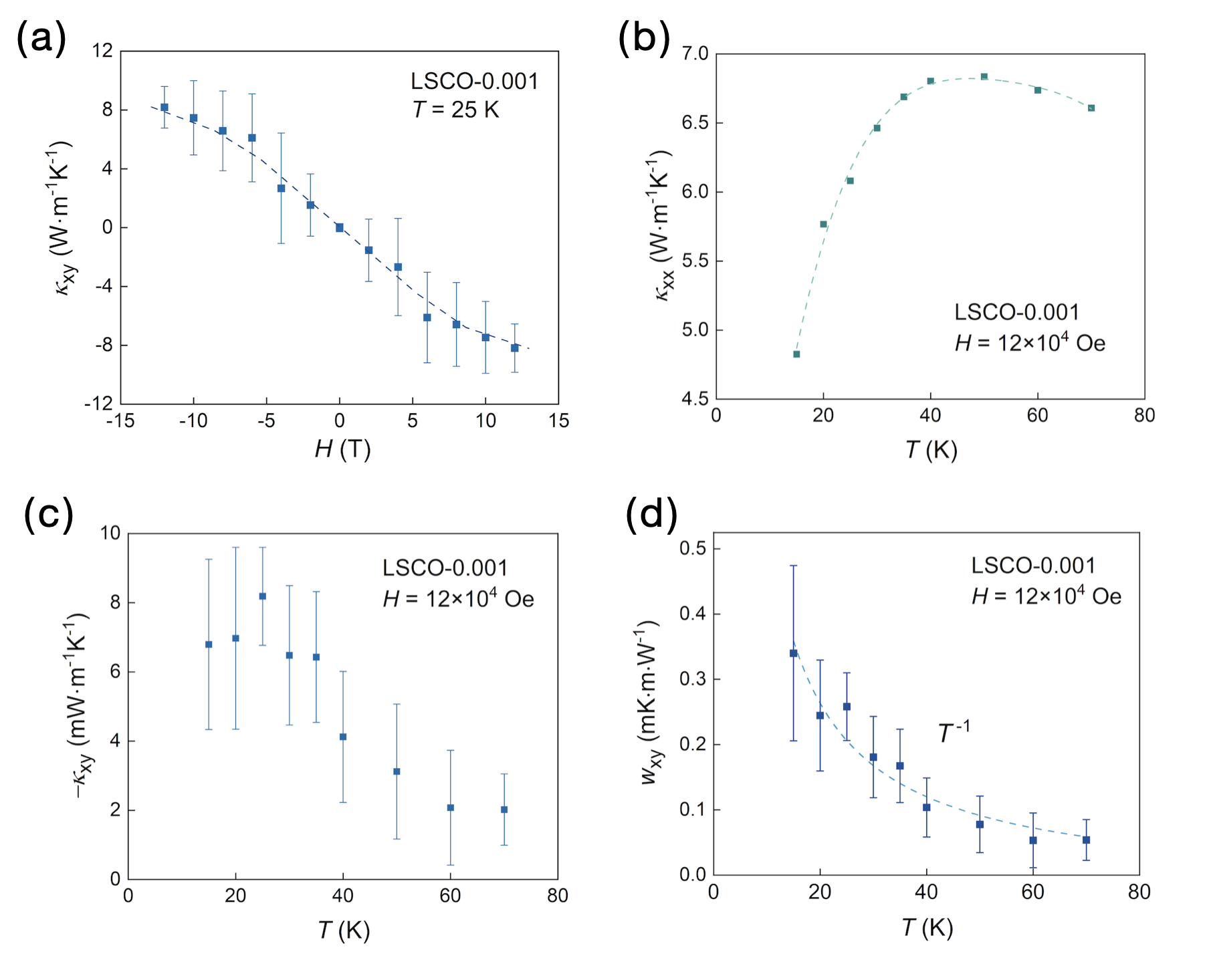}
	}       
	\caption{(a) Field dependence of the thermal Hall conductivity $\kappa_{xy}$ of LSCO-0.001 at 25 K. Temperature dependence of (b) longitudinal thermal conductivity, (c) thermal Hall conductivity, and (d) thermal Hall resistivity of LSCO-0.001 at \mbox{$12 \times 10^4$~Oe}. The thermal Hall resistivity displays a $T^{-1}$-divergent behavior at low temperature.
		\label{fig:S9}
	}
\end{figure}

\LSCOthousandth\ (LSCO-0.001) and \LSNOthird\ (LSNO-1/3) are both doped Mott insulators with similar layered crystal structure, yet they differ in their magnetic and electronic ground states. LSCO-0.001 exhibits three-dimensional long-range antiferromagnetic (AFM) order without stripe formation, while LSNO-1/3 hosts static charge and spin ordered stripes with quasi-one-dimensional antiferromagnetic (AFM) correlations at low temperatures~\cite{Viskadourakis2015Cuprates, 58_Freeman2011}.

We measured the thermal transport in LSCO-0.001 using a thermocouple setup in the Physical Property Measurement System (PPMS). Figure~\ref{fig:S9}(a) shows the field dependence of $\kappa_{xy}$ of LSCO-0.001 at 25 K. The $\kappa_{xy}(H)$ exhibits sub-linear behavior, similar to the $\kappa_{xy}(H)$ of LSNO-1/3 [Fig.~1(b) in the main text]. Figure~\ref{fig:S9}(b) and (c) show the temperature dependence of $\kappa_{xx}$ and $\kappa_{xy}$, respectively, of LSCO-0.001 at $12\times10^4$~Oe. There exists a peak in the $\kappa_{xx}(T)$, similar to the $\kappa_{xx}(T)$ of LSNO-1/3 [Fig.~1(c) in the main text]. Figures \ref{fig:S9}(d) presents the temperature dependence of the thermal Hall resistivity $w_{xy}$, which follows a $T^{-1}$ behavior for the entire measured range of 15--70~K. This behaviour is consistent with the $w_{xy}(T)$ of LSNO-1/3 [Fig.~2(a) in the main text] and the $w_{xy}(T)$ of La$_2$CuO$_4$ [Fig.~3(b) in the main text]. 

The thermal Hall data of LSCO-0.001 resemble that of LSNO-1/3 as discussed in the main text. The consistent results obtained from two independent measurement techniques also provide mutual validation of the data reliability and further support the unifying phenomena discussed in the main text.
\newpage

\section{S10. Extended thermal Hall measurement of LSNO-1/3 at higher temperatures}

\begin{figure}[h]\centering
	\resizebox{18cm}{!}{
		\includegraphics{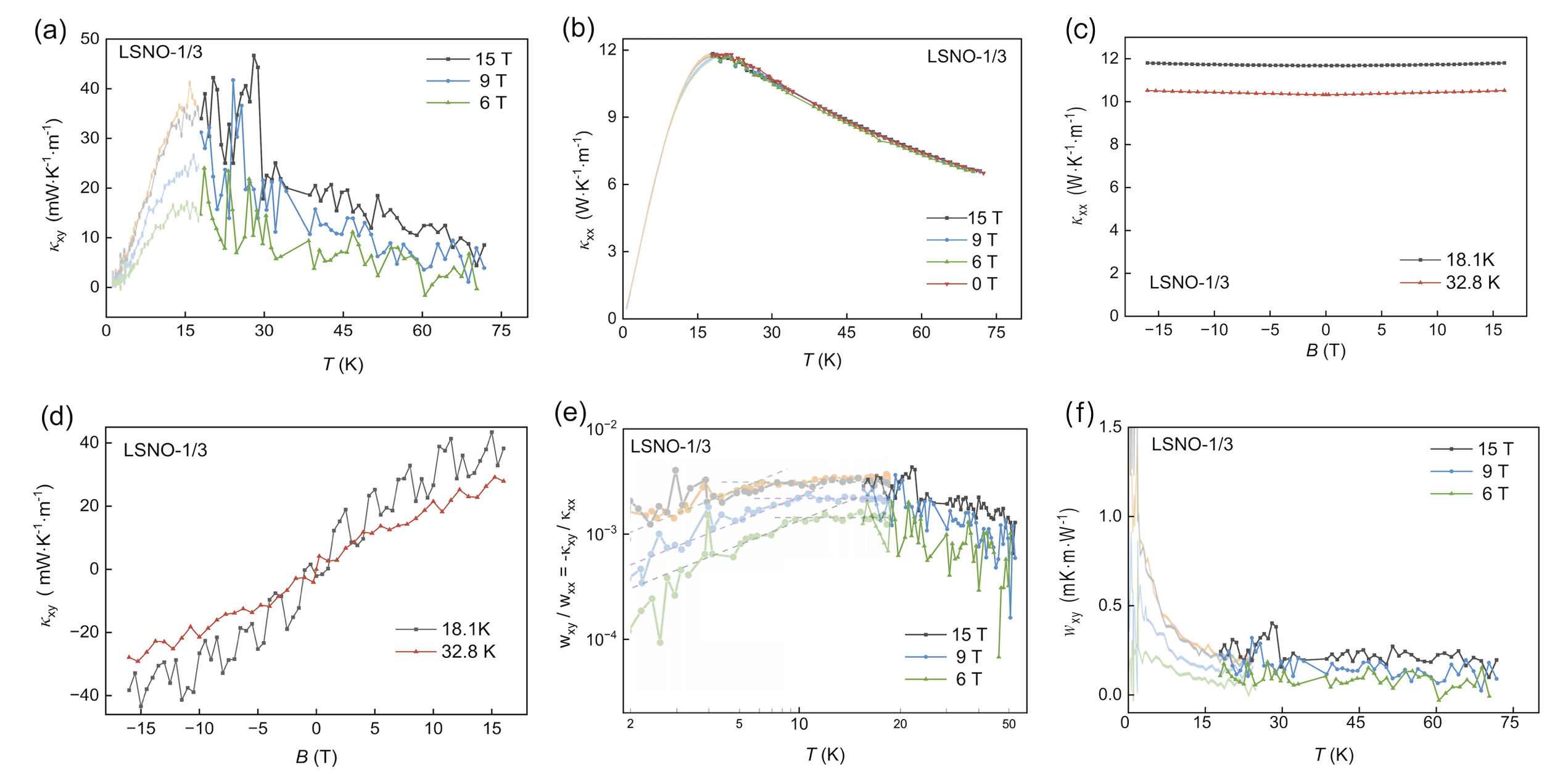}
	}       
	\caption{ 
		Temperature dependence of (a)~thermal Hall conductivity $\kappa_{xy}$ and (b)~longitudinal thermal conductivity $\kappa_{xx}$ of LSNO-1/3 in various magnetic fields. Field dependence of (c)~$\kappa_{xx}$ and (d)~$\kappa_{xy}$ at 18.1 K and 32.8 K. Temperature dependence of (e)~the thermal Hall angle and (f)~the thermal Hall resistivity $w_{xy}$ of LSNO-1/3 in different magnetic fields. The light-colored traces represent the low-temperature data presented in the main text, while the dark-colored curves show higher-temperature data measured in a PPMS.
		\label{fig:S10}
	}
\end{figure}

To extend the thermal Hall characterization of \LSNOthird\ to higher temperatures, complementary measurements were performed in a PPMS system on the same \LSNOthird\ sample, using Type-K thermocouples to detect the transverse temperature difference while $\kappa_{xx}$ was measured with Cernox thermometers. Figure~\ref{fig:S10} summarizes the results of thermal conductivity and thermal Hall measurements of \LSNOthird\ across a wide temperature range. The transparent background traces below $\sim$20~K in panels (a, b, e, f) represent the low-temperature data presented in the main text.

Figure~\ref{fig:S10}(a) shows the temperature dependence of $\kappa_{xy}$, which peaks near 20~K and gradually decreases as temperature increases. The temperature dependence of $\kappa_{xx}$ is shown in Fig.~\ref{fig:S10}(b) at the same fields, exhibiting a similar peak around 20~K followed by a gradual decrease. The results from 1~K to 70~K indicate clearly the longitudinal and transverse thermal responses share a common intermediate-temperature peak feature as identified in the main text.

To further study the field dependence at high temperatures, Fig.~\ref{fig:S10}(c) and (d) plot the field dependence of $\kappa_{xx}$ and $\kappa_{xy}$ at $T = 18.1$~K and 32.8~K, respectively. $\kappa_{xx}$ varies slightly with field at both temperatures, indicating a weak magneto-thermal conductivity. $\kappa_{xy}$ exhibits a sub-linear field dependence at both temperatures, consistent with the field dependence of $\kappa_{xy}$ at low temperatures reported in main text.

The temperature dependence of the thermal Hall angle $w_{xy}/w_{xx}$ is shown in Fig.~\ref{fig:S10}(e) on a log-log scale at different fields. The thermal Hall angle increases to a peak around 20~K and decreases gradually at higher temperatures. Figure~\ref{fig:S10}(f) shows the temperature dependence of $w_{xy}$, which connects smoothly to the low-temperature data.

Despite larger noise, the high-temperature measurement are in good agreement with the low-temperature measurement presented in the main text. This validates the two independent measurement techniques, supporting the reliability of our thermal Hall experiment of \LSNOthird\ across the full measured temperature range.
\newpage

\providecommand{\noopsort}[1]{}\providecommand{\singleletter}[1]{#1}%

\end{document}